\documentclass[aps, pra, superscriptaddress, notitlepage, reprint]{revtex4-2}
\usepackage[english]{babel}
\usepackage{amsmath,amsthm}
\usepackage{amsfonts}
\usepackage[pdfborder={0 0 0}, colorlinks=true, urlcolor=blue, linkcolor=blue, citecolor=blue]{hyperref}
\usepackage{color}
\usepackage{graphicx}
\usepackage{float}
\usepackage{subfigure}
\usepackage{lipsum}
\usepackage[T1]{fontenc}

\theoremstyle{definition}

\theoremstyle{remark}

\begin{document}

\title{Phase-controlled bipartite and tripartite entanglement and Bell nonlocality in a closed-loop optomechanical system}
\author{Sha-Sha Zheng}
\affiliation{Beijing Institute of Radio Metrology and Measurement, Beijing 100854, China}
\affiliation{Science and Technology on Metrology and Calibration Laboratory, Beijing Institute of Radio Metrology and Measurement, Beijing 100854, China}
\author{Zi-Nan Wu}
\affiliation{Beijing Institute of Radio Metrology and Measurement, Beijing 100854, China}
\affiliation{Science and Technology on Metrology and Calibration Laboratory, Beijing Institute of Radio Metrology and Measurement, Beijing 100854, China}
\author{Di-Di Xu}
\affiliation{Beijing Institute of Radio Metrology and Measurement, Beijing 100854, China}
\affiliation{Science and Technology on Metrology and Calibration Laboratory, Beijing Institute of Radio Metrology and Measurement, Beijing 100854, China}
\author{Chang-Yu Li}
\affiliation{Beijing Institute of Radio Metrology and Measurement, Beijing 100854, China}
\affiliation{Science and Technology on Metrology and Calibration Laboratory, Beijing Institute of Radio Metrology and Measurement, Beijing 100854, China}
\author{Peng-Wei Gong}
\email[Electronic address: ]{pwgong_birmm@163.com}
\affiliation{Beijing Institute of Radio Metrology and Measurement, Beijing 100854, China}
\affiliation{Science and Technology on Metrology and Calibration Laboratory, Beijing Institute of Radio Metrology and Measurement, Beijing 100854, China}
\author{Qiongyi He}
\affiliation{State Key Laboratory of Artificial Microstructure and Mesoscopic Physics, School of Physics, Frontiers Science Center for Nano-optoelectronics, $\&$ Collaborative Innovation Center of Quantum Matter, Peking University, Beijing 100871, China}
\affiliation{Collaborative Innovation Center of Extreme Optics, Shanxi University, Taiyuan 030006, China}
\affiliation{Hefei National Laboratory, Hefei 230088, China}
\author{Feng-Xiao Sun}
\email[Electronic address: ]{sunfengxiao@bupt.edu.cn}
\affiliation{State Key Laboratory of Information Photonics and Optical Communications and School of Physical Science and Technology, Beijing University of Posts and Telecommunications, Beijing 100876, China}

\begin{abstract}
We propose a novel scheme to generate and manipulate bipartite and tripartite entanglement and Bell nonlocality in a closed-loop three-mode optomechanical system, where two optical modes are simultaneously coupled to a mechanical mode via typical optomechanical interactions and also coupled to each other through field transmission. This configuration gives rise to a phase-sensitive coupling, in which the relative phase directly controls the population distribution between two optical modes, enabling coherent redistribution of quantum correlations. By tuning this relative phase, we achieve deterministic switching of bipartite entanglement between two optical-mechanical pairs, as well as tunable genuine tripartite entanglement among three modes. Employing the displaced-parity measurement, we construct both the bipartite and tripartite Bell inequalities in phase space and observe maximal violations of approximately $2.32$ and $3$, respectively, which coincide exactly with the values achievable for ideal bipartite and tripartite Einstein-Podolsky-Rosen states. Counterintuitively, we find that the tripartite Bell nonlocality can persist even when the genuine tripartite entanglement is absent, providing deeper insight into the relationship between these two types of quantum correlations.
Furthermore, we systematically analyze the effects of mechanical dissipation and thermal noise, identifying the parameter regions where Bell violation survives under realistic experimental conditions.
Our results establish a comprehensive framework for generating, controlling, and verifying multipartite quantum correlations, paving the way towards phase-tunable quantum networks and noise-resilient tests of quantum foundations.
\end{abstract}
\maketitle

\section{Introduction}
Quantum entanglement~\cite{Einstein1935,Schrodinger1935} and Bell nonlocality~\cite{Bell1964} represent two fundamental features of quantum mechanics that distinguish it from classical physics. Entanglement characterizes non-classical correlations between subsystems that cannot be reproduced by separable states, whereas Bell nonlocality refers to correlations incompatible with local hidden-variable models~\cite{Horodecki2009,Brunner2014RMP,Uola2020RMP}. Although these two quantum correlations are equivalent for pure quantum states~\cite{Yu2012All}, their relationship becomes profoundly more intricate in mixed states, where entanglement does not necessarily imply Bell nonlocality~\cite{Werner1989PRA}. Accordingly, Bell nonlocality constitutes a strict subset of entanglement~\cite{Wiseman2007, Jones2007,Cavalcanti2009}. However, in multipartite scenarios, the complexity is further amplified as the exact hierarchy between genuine multipartite entanglement and multipartite Bell nonlocality remains far from fully elucidated~\cite{Brunner2014RMP}. 

The preparation and manipulation of quantum correlations in multipartite systems at mesoscopic and macroscopic scales are of great importance for both fundamental studies and quantum technologies. For instance, quantum entanglement and Bell nonlocality serve as powerful tools for investigating fundamental issues in quantum mechanics, including decoherence~\cite{MarkusAspelmeyer2012OptomechanicsDecoherence}, quantum-to-classical crossover~\cite{FrancoNori2006QuantumClassicalTransition,TonyELee2013QuantumClassicalTransition}, and wave function collapse models~\cite{Angelo2013RMP_WaveFunctionCollapse}. From the perspective of practical applications, quantum entanglement is an indispensable resource for quantum communication~\cite{Bennett1993,Gisin2002,Simon2017QuantumCommunication}, quantum sensing~\cite{Degen2017QuantumSensing,Pirandola2018QuantumSensing}, quantum computation~\cite{Nielsen2010QuantumComputation}, and quantum simulation~\cite{Georgescu2014QuantumSimulation}. Bell nonlocality, in turn, underpins device-independent quantum-information tasks, including secure quantum key distribution~\cite{Acin2007PRL} and certified randomness generation~\cite{Pironio2010Nature,Li2025Necessary}. Thus, understanding and controlling multipartite quantum correlations in mesoscopic and macroscopic systems represents a crucial stepping stone for both advancing foundational quantum inquiries and enabling practical quantum technologies.

Cavity optomechanical systems~\cite{Markus2014RMP,PhysicsToday2012,Florian2009Review,MetcalfeMichael2014OptomechanicsReview}, which govern the coherent coupling between the electromagnetic field and the mechanical degrees of freedom mediated by the radiation pressure force, have emerged as a highly promising platform for generating and manipulating quantum effects at the mesoscopic and macroscopic scales~\cite{Pirkkalainen2015MechanicalSqueezing,Wollman2015MechanicalSqueezing,Lecocq2015MechanicalSqueezing,HuangXinyao2018Mech-MechEnt,Thomas2021Mech-MechEnt,Kotler2021Mech-MechEnt,Mercier2021Mech-MechEnt,Clerk2021Mech-MechEnt-Perspective,YuHaocun2020LIGO_QuantumCorrelation,Palomaki710,Vivoli2016PRL_OptBell,Hofer2016PRL_OptBell,Riedinger2018PCcavity_Ent,Ockeloen-Korppi2018Mech-MechEnt,Marinkovic2018PRL_OptBell}.
Considerable theoretical and experimental effort has been devoted to generating and controlling bipartite entanglement and Bell nonlocality in optomechanical systems~\cite{Vitali2007PRL_OptoEnt,Genes2008NJP_Ent,Barzanjeh2011PRA_Ent,Riedinger2018PCcavity_Ent,Ockeloen-Korppi2018Mech-MechEnt,Palomaki710,Vivoli2016PRL_OptBell,Hofer2016PRL_OptBell,Marinkovic2018PRL_OptBell}. Although various theoretical schemes have extended the generation of optomechanical entanglement to multipartite settings~\cite{Xuereb2012PRA_MultiEnt,WangYingdan2015PRA_MultiEnt,XiangYu2015OE_MultiEnt,FrancoNori2022PRA_MultiEnt,JingHui2025PRA_MultiEnt}, investigations of Bell nonlocality in these systems have remained predominantly confined to bipartite configurations~\cite{LiJie_2017PRA_OptoBell,Muhammad2018PRA_OptBell,ZhangJing2015AnnPhys_OptBell,TanHuatang2020QST_OptBell}. Consequently, the systematic investigation of multipartite Bell nonlocality and its relation to genuine multipartite entanglement within a single optomechanical device remains largely unexplored.

\begin{figure}
\centering
\includegraphics[width=1.0\columnwidth]{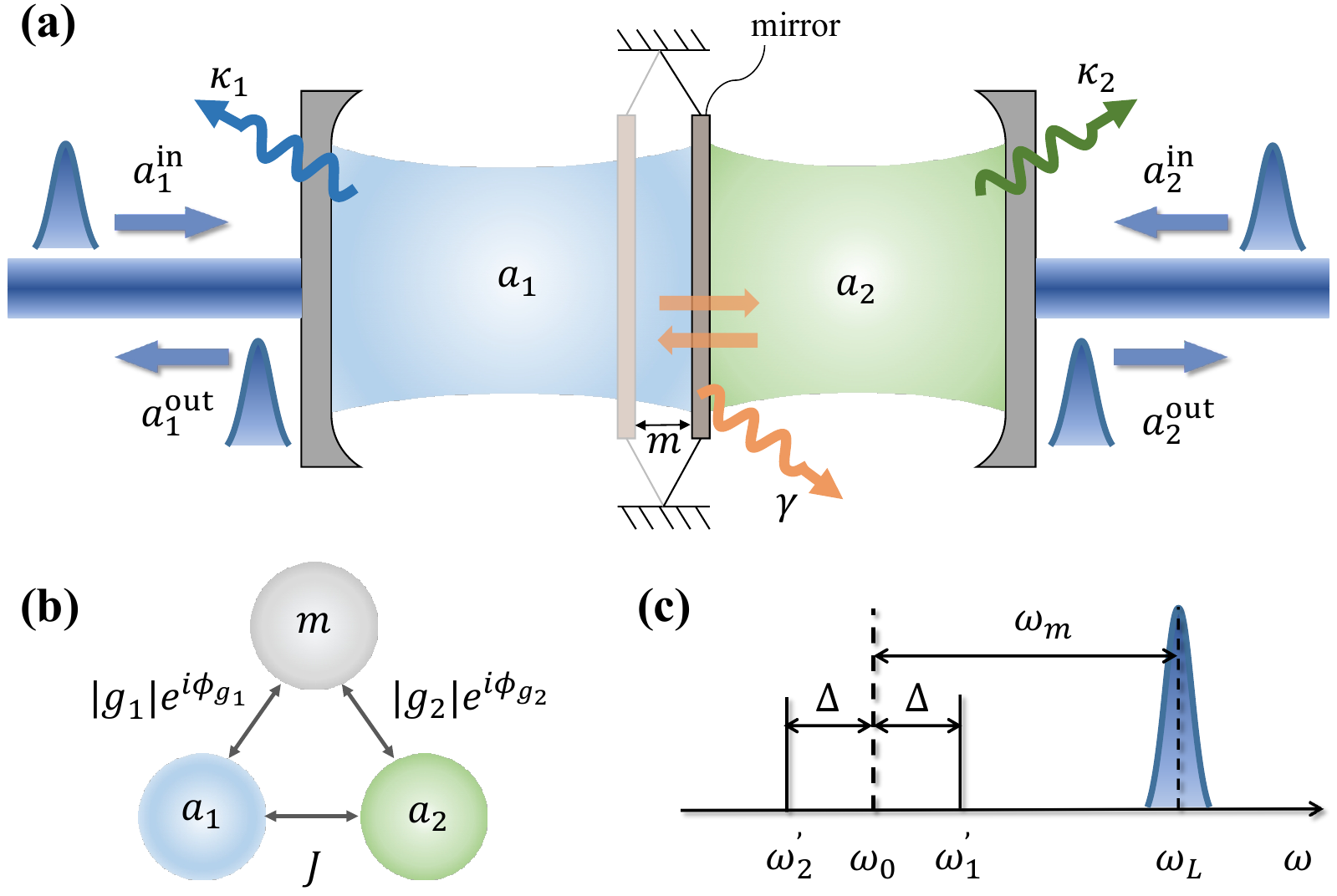}
\caption{Schematic diagram of the closed-loop three-mode cavity optomechanical system. (a) The system consists of a partially transmitting dielectric membrane placed in the middle of an optical cavity, which splits the cavity into two coupled modes described by annihilation operators $a_1$ and $a_2$ with frequencies $\omega_1$ and $\omega_2$, respectively, and a mechanical mode represented by $m$ with frequency $\omega_m$. The two cavity modes are driven by a pulsed laser of duration time $\tau$ and are damped with rates $\kappa_1$ and $\kappa_2$. The mechanical mode is damped with rate $\gamma$. The input and output fields for the cavity modes are denoted by $a_{1,2}^{\rm in}$ and $a_{1,2}^{\rm out}$, respectively.
(b) The simplified interaction diagram of the close-loop three-mode system. Here $g_{1}=|g_{1}|\exp(i\phi_{g_1})$ and $g_{2}=|g_{2}|\exp(i\phi_{g_2})$ respectively represent the effective optomechanical coupling strengths between the two cavity modes and the mechanical mode, while $J$ denotes the linear coupling strength between the two cavity modes.
(c) The laser frequency is tuned to $\omega_L = \omega_0 + \omega_m$, i.e., the blue (anti-Stokes) sideband of the average cavity frequency $\omega_0 = (\omega_1' + \omega_2')/2$, where $\omega_{j}' = \omega_j + g_{0,j}(\beta + \beta^*)$ denotes the effective frequency of each cavity mode.}
\label{Fig_Scheme}
\end{figure}

Motivated by the above insights, we investigate a closed-loop three-mode optomechanical system, in which two optical modes and one mechanical mode are mutually coupled, forming a cyclic and phase-sensitive configuration~\cite{Sun_2017NJP}.
The relative phase of these couplings determines the population distribution of the optical modes and coherently redistributes quantum correlations among the three modes, enabling deterministic switching of bipartite entanglement and Bell nonlocality between the two optical-mechanical pairs. Moreover, genuine tripartite entanglement can be generated and optimized, with the tripartite Bell nonlocality manifested over a wide parameter range. A counterintuitive finding reveals that tripartite Bell nonlocality can persist even when the genuine tripartite entanglement vanishes, demonstrating that genuine tripartite entanglement is not a necessary condition for tripartite Bell nonlocality. We further examine the effects of mechanical dissipation and thermal noise and identify parameter regimes in which Bell violations remain observable under experimentally realistic conditions. These results establish the proposed system as a versatile platform for phase-controlled multipartite quantum entanglement and Bell nonlocality.

The structure of this paper is arranged as follows. In Sec.~\ref{Model}, we  introduce the theoretical model of the proposed closed-loop tripartite optomechanical system, consisting of presenting the Hamiltonian and deriving the corresponding equations of motion. The covariance matrix is then analytically constructed, providing the essential framework for analyzing bipartite and tripartite quantum correlations.
Then in Sec.~\ref{Criteria}, we recall the criteria adopted in this paper to test both bipartite and tripartite quantum entanglement and Bell nonlocality. In Sec.~\ref{NumericalResults}, we numerically simulate the phase control effects of both bipartite and tripartite entanglement as well as Bell nonlocality. 
Besides, the effects of the squeezing parameter, the mechanical dissipation and thermal noise on relevant quantum correlations are explored. Finally, the content of this paper is summarized in Sec.~\ref{Conclusion}.

\section{Model}\label{Model}
\subsection{System Hamiltonian and Equations of Motion}

We consider a closed-loop three-mode cavity optomechanical system, as illustrated in Fig.~\ref{Fig_Scheme}, composed of two cavity field modes and one mechanical mode. Two cavity modes (represented by annihilation operators $a_{1}$ and $a_{2}$) interact with each other through field transmission, and simultaneously interact with the mechanical mode (represented by the annihilation operator $m$) through typical optomechanical interactions.
The Hamiltonian of the proposed system takes the form of 
\begin{eqnarray}
H &=& \hbar\omega_{1}a^{\dag}_{1}a_{1}+\hbar\omega_{2}a^{\dag}_{2}a_{2}+\hbar\omega_{m}m^{\dag}m+\hbar J(a_{1}^{\dag}a_{2}+ a_{2}^{\dag}a_{1}) \nonumber \\
&&+\hbar (g_{0,1}a^{\dag}_{1}a_{1} + g_{0,2}a^{\dag}_{2}a_{2})(m^{\dag}+m) \nonumber\\
&&+i\hbar[E_{1}(t)a_{1}^{\dag} +E_{2}(t)a_{2}^{\dag} - {\rm H.c.}] ,
\label{Hamiltonian}
\end{eqnarray}
where $\omega_{1},~\omega_{2}$ and $\omega_{m}$ respectively  represent the frequency of two cavity modes $a_{1},~a_{2}$ and the mechanical mode $m$. $J$ denotes the direct linear coupling strength between two cavity modes, while $g_{0,j}$ is the single-photon coupling strength of the cavity mode $a_{j}$ to the mechanical mode $m$.
Two cavity modes are respectively driven by short laser pulses with the same driving frequency $\omega_{L}$ and duration time $\tau$, but with different driving amplitude (denoted as $E_{01}$ and $E_{02}$) and phase (denoted as $\varphi_{01}$ and $\varphi_{02}$). Therefore, the driving term can be well described by the last term of 
Eq.~(\ref{Hamiltonian}), where electric field amplitudes read $E_{1}(t)=E_{01}\exp[-i(\omega_{L}t +\varphi_{01})]$ and $E_{2}(t)=E_{02}\exp[-i(\omega_{L}t +\varphi_{02})]$. The emergence of pairwise coupling among three modes creates a closed-loop three-mode optomechanical system, as illustrated in Fig.~\ref{Fig_Scheme}(b). This cyclic topology induces phase‑sensitive interference effects that fundamentally govern the dynamical evolution of the constituent modes.
The above closed-loop system can be experimentally realized in a typical optomechanical system composed of a membrane in a Fabry-P\'erot cavity~\cite{Markus2014RMP,Sun_2017NJP}. 

Substituting the Hamiltonian (\ref{Hamiltonian}) into the Heisenberg equation of motion and incorporating dissipation through the quantum noise approach, the quantum Langevin equations in the rotating frames relative to the driving laser (i.e., $a_{j}\rightarrow a_{j}\exp\left[i\left( \omega_{L}t+\varphi_{0j}\right)\right], j=1,~2$) can be derived as
\begin{eqnarray} 
	\dot{a}_{1}&= & -(\kappa_{1}+i\Delta_{1})a_{1}-ig_{0,1}a_{1}(m+m^{\dagger}) -iJa_{2}e^{i\delta\varphi_0}\nonumber\\
&&+E_{01}-\sqrt{2\kappa_{1}}a_{1}^{{\rm in}} , \nonumber\\
	\dot{a}_{2}&=& -(\kappa_{2}+i\Delta_{2})a_{2}-ig_{0,2}a_{2}(m+m^{\dagger})-iJa_{1}e^{-i\delta\varphi_0}\nonumber\\
&&+E_{02}-\sqrt{2\kappa_{2}}a_{2}^{{\rm in}} , \nonumber\\
	\dot{m} &=& -(\gamma+i\omega_{m})m-ig_{0,1}a_{1}^{\dagger}a_{1}-ig_{0,2}a_{2}^{\dagger}a_{2}-\sqrt{2\gamma}m^{\rm in} .\nonumber\\
\label{Langevin0}
\end{eqnarray} 
Here, $\Delta_{j}=\omega_{j}-\omega_{L}$ denotes the detuning of the cavity modes $a_j$ from the laser frequency. $\kappa_{j}$ and $\gamma$ correspond to the damping rates of the cavity mode $a_j$ and mechanical mode $m$, respectively. $\delta\varphi_0=\varphi_{01}-\varphi_{02}$ represents the phase difference. $a^{\rm in}_{j}$ and $m^{\rm in}$ denote the input quantum noises, which arise from the inevitable coupling of the modes to their surrounding environments. Generally, the quantum noise of the mechanical mode is assumed to reside in a thermal state characterized by the second-order correlation functions $\langle m^{\rm in}(t)(m^{\rm in})^{\dagger}(t')\rangle = \left(n_m+1\right)\delta(t-t')$ and $\langle (m^{\rm in})^{\dag}(t)m^{\rm in}(t')\rangle = n_m\delta(t-t')$ with zero mean value, where $n_m=1/[\exp(\hbar\omega_{m}/k_{B}T)-1]$ refers to the average thermal phonon number of the mechanical mode, $k_B$ denotes the Boltzmann constant and $T$ represents the temperature of the surrounding environment. Correspondingly, the quantum noise of the optical mode is in a vacuum state characterized by the second-order correlation function $\langle a^{\rm in}_{j}(t)(a_{j}^{\rm in})^{\dagger}(t')\rangle =  \delta(t-t')$ due to its high frequency.

Under strong laser driving where the intracavity photon number is large and the steady-state classical average values of the optical fields satisfy $\langle a_j\rangle \gg 1$, the standard linearization approach can be applied~\cite{Vitali2007PRL_OptoEnt,Vitali2007JPA_Ent,Genes2008NJP_Ent,Genes2009BookChapter,Barzanjeh2011PRA_Ent,braunstein2012quantum}. This well-established method, widely used in optomechanics~\cite{Markus2014RMP,Marquardt2007CoolingLimit,Wilson-Rae2007CoolingLimit,LiuYongChun_2013Review_Cooling}, decomposes each field annihilation operator into its steady-state classical mean value and a small quantum fluctuation around the steady state, namely $O = \langle O\rangle + \delta O, \quad O \in \{a_1, a_2, m, a_1^{\mathrm{in}}, a_2^{\mathrm{in}}, m^{\mathrm{in}}\}$.
Substituting this expansion into the Langevin equations~(\ref{Langevin0}) and assuming $\langle O\rangle \gg \delta O$, the equations for the steady-state classical averages are obtained by setting $\dot{\langle O\rangle}=0$ and neglecting all quantum fluctuation terms, which yields
\begin{eqnarray} 
0&= & -\left(\kappa_{1}+i\Delta_{1}\right)\alpha_{1}-ig_{0,1}\alpha_{1}\left(\beta+\beta^*\right) -iJ\alpha_{2}e^{i\delta\varphi_0}\nonumber\\
&&+E_{01}, \nonumber\\
0&=& -\left(\kappa_{2}+i\Delta_{2}\right)\alpha_{2}-ig_{0,2}\alpha_{2}\left(\beta+\beta^*\right)-iJ\alpha_{1}e^{-i\delta\varphi_0}\nonumber\\
&&+E_{02}, \nonumber\\
0&=& -\left(\gamma+i\omega_{m}\right)\beta-ig_{0,1}\left|\alpha_{1}\right|^2-ig_{0,2}\left|\alpha_{2}\right|^2.
\label{Langevin0_MeanValue0}
\end{eqnarray} 
Here, we denote $\langle a_j\rangle = \alpha_j\;(j=1,~2)$ and $\langle m\rangle = \beta$. The solution of the above equations takes the form of 
\begin{eqnarray}
\alpha_{1} &=& \frac{(\kappa_{2}+i\Delta'_{2})E_{01}e^{i\varphi_{01}} -iJE_{02}e^{i\varphi_{02}}}{(\kappa_{1}+i\Delta'_{1})(\kappa_{2}+i\Delta'_{2})+J^{2}}, \nonumber\\
\alpha_{2} &=& \frac{(\kappa_{1}+i\Delta'_{1})E_{02}e^{i\varphi_{02}} -iJE_{01}e^{i\varphi_{01}}}{(\kappa_{1}+i\Delta'_{1})(\kappa_{2}+i\Delta'_{2})+J^{2}} , \nonumber\\
\beta &=& \frac{-i(g_{0,1}|\alpha_1|^2 +g_{0,2}|\alpha_2|^2)}{\gamma+i\omega_m} ,\label{Langevin0_MeanValue}
\end{eqnarray}
where the effective detuning reads $\Delta'_{j} = \omega'_{j} -\omega_L$ with $\omega'_{j} = \omega_{j}+ g_{0,j}(\beta+\beta^{\ast}) $.

Once the steady-state solutions are obtained, the linearized quantum Langevin equations for the quantum fluctuation operators $\delta O$ can be readily derived. Neglecting the high-order terms and retaining only the linear contributions, the quantum Langevin equations can be formulated as
\begin{eqnarray}
	\delta \dot{a}_{1}&=& -(\kappa_{1}+i\Delta'_{1})\delta a_{1}-ig_1e^{i\varphi_{01}}(\delta m+\delta m^\dagger)\nonumber\\
&&-iJe^{i\delta\varphi_0}\delta a_{2}-\sqrt{2\kappa_{1}}\delta a^{\rm in}_{1}, \nonumber\\
\delta \dot{a}_{2}&=& -(\kappa_{2}+i\Delta'_{2})\delta a_{2}-ig_2e^{i\varphi_{02}}(\delta m+\delta m^\dagger)\nonumber\\
&&-iJe^{-i\delta\varphi_0}\delta a_{1}-\sqrt{2\kappa_{2}}\delta a^{\rm in}_{2}, \nonumber\\
\delta \dot{m} &=& -(\gamma+i\omega_m )\delta m -i(g_1^{*}e^{-i\varphi_{01}}\delta a_1+g_1e^{i\varphi_{01}}\delta a_1^\dagger) \nonumber\\
&&-i(g_2^*e^{-i\varphi_{02}}\delta a_2+g_2e^{i\varphi_{02}}\delta a_2^\dagger)-\sqrt{2\gamma}\delta m^{\rm in} ,
\label{Langevin0_QuantumFluctuations0}
\end{eqnarray}
with $g_j=g_{0,j}\alpha_j e^{-\varphi_{0j}}$ denoting the effective optomechanical coupling strength between the cavity mode $a_j$ and the mechanical mode $m$. Without loss of generality, we hereafter express the coupling as $g_{1}=|g_{1}|\exp(i\phi_{g_1})$ and $g_{2}=|g_{2}|\exp(i\phi_{g_2})$.

Equation~(\ref{Langevin0_QuantumFluctuations0}) reveals that the quantum fluctuation operators oscillate at frequencies $\pm\omega_{m}$. 
Therefore, it is convenient to introduce slowly varying fluctuation operators $\delta m\rightarrow  \delta m e^{i\omega_mt},\ \delta m^{\rm in}\rightarrow  \delta m^{\rm in}e^{i\omega_mt}, \ \delta a_{j} \rightarrow  \delta a_{j}e^{-i(\omega_mt +\varphi_{0j})}$, $\delta a^{\rm in}_{j} \rightarrow  \delta a^{\rm in}_{j}e^{-i(\omega_mt +\varphi_{0j})}$. Substituting these slowly varying fluctuation operators into Eq.~(\ref{Langevin0_QuantumFluctuations0}) and applying the rotating-wave approximation (neglecting all terms oscillating at $2\omega_m$), we obtain
\begin{eqnarray}
\dot{a}_{1} &=& -\left(\kappa_{1}+i\Delta\right)a_{1}-ig_1m^{\dag}-iJa_{2}-\sqrt{2\kappa_{1}}a^{\rm in}_{1} ,\nonumber\\
\dot{a}_{2} &=& -\left(\kappa_{2}-i\Delta\right)a_{2}-ig_2m^{\dag}-iJa_{1}-\sqrt{2\kappa_{2}}a^{\rm in}_{2} ,\nonumber\\
\dot{m} &=& -\gamma m-ig_1a_{1}^{\dag}-ig_2a_2^\dag-\sqrt{2\gamma}m^{\rm in} .
\label{Langevin0_QuantumFluctuations}
\end{eqnarray}
Here, as shown in Fig.~\ref{Fig_Scheme}(c), we have assumed the laser pulses are tuned to the blue sideband of the mean cavity frequency, i.e., $\omega_L = \omega_0 + \omega_m$ with $\omega_0 = (\omega_1' + \omega_2')/2$ and $\Delta = (\omega_1' - \omega_2')/2$. For clarity of the notation, the symbol $\delta$ has been omitted in Eq.~(\ref{Langevin0_QuantumFluctuations}).

\subsection{Derivation of the Covariance Matrix}
Having introduced the system Hamiltonian and systematically derived the equations of motion for the proposed closed‑loop three‑mode optomechanical system in the preceding subsection, we now proceed to construct the corresponding covariance matrix, which provides the essential framework for characterizing both entanglement and Bell nonlocality within the system.

To simplify the calculation and gain deeper physical insight, we introduce orthogonal superposition modes
\begin{eqnarray}
a_{w} &=& a_{1}\cos\theta +a_{2}\sin\theta ,\nonumber \\
a_{u} & =& a_{1}\sin\theta  -a_{2}\cos\theta ,
\label{OrthogonalSuperpositionModes}
\end{eqnarray} 
where $\cos^2\theta =1/2+\Delta/(2w)$ with $w=\sqrt{J^{2}+\Delta^{2}}$. The angle $\theta$ belongs to the interval $[0,\pi/2]$. 
In the basis of the superposition modes, the equations of motion given in Eq.~(\ref{Langevin0_QuantumFluctuations}) simplify to
\begin{eqnarray}
\dot{a}_{w} &=&-(\kappa +iw)a_{w}-ig_{w}m^{\dag}-\sqrt{2\kappa}a^{\rm in}_{w} ,\nonumber\\
\dot{a}_{u} &=&-(\kappa -iw)a_{u}-ig_{u}m^{\dag}-\sqrt{2\kappa}a^{\rm in}_{u} ,\nonumber\\
\dot{m} &=&-\gamma m -i\left(g_{w}a^\dagger_{w}+g_{u}a^\dagger_{u}\right) -\sqrt{2\gamma}m^{\rm in}.
\label{Langevin0_QuantumFluctuations_SuperPosition}
\end{eqnarray}
The coefficients $g_{w} = g_{1}\cos\theta + g_{2}\sin\theta$ and  $g_{u} = g_{1}\sin\theta - g_{2}\cos\theta $, which are respectively the effective coupling strengths between the superposition modes $a_w$ and $a_u$ with the mechanical mode $m$. In the above derivation, we have assumed identical dissipation rates for the two cavity modes, i.e., $\kappa_1 = \kappa_2 = \kappa$.

Obtaining a general solution of Eq.~(\ref{Langevin0_QuantumFluctuations_SuperPosition}) is especially challenging. However, a simple analytical form emerges in two limiting cases: the bad‑cavity regime, where $\kappa \gg |g_{w,u}|$, or when the frequency splitting between the superposition modes is large, i.e., $w\gg |g_{w,u}|$. In either case, the superposition modes $a_{w}$ and $a_{u}$ evolve slowly, allowing the adiabatic approximations $\dot{a}_{w}\approx 0$ and $\dot{a}_{u}\approx 0$. This yields
\begin{eqnarray}
a_{w} &=& \frac{-e^{-i\phi}}{\sqrt{\kappa^2+w^2}}(ig_w m^\dagger+\sqrt{2\kappa}a^{\rm in}_w),\nonumber \\
a_{u}  &=& \frac{-e^{i\phi}}{\sqrt{\kappa^2+w^2}}(ig_u m^\dagger+\sqrt{2\kappa}a^{\rm in}_u) ,\label{AdiabaticApprox_OrthogonalSuperpositionModes}
\end{eqnarray}
where $\phi=\arctan \left(w/\kappa\right)$. Therefore, the equation of motion for the mechanical mode becomes
\begin{equation}
\dot{m} = (G+i\delta)m+i\sqrt{2G_w}e^{i\phi}a_w^{\rm in\dagger}+i\sqrt{2G_u}e^{-i\phi}a_u^{\rm in\dagger}-\sqrt{2\gamma}m^{\rm in} ,
\label{AdiabaticApprox_MechanicalMode}
\end{equation}
with $\delta = (|g_w|^2 -|g_u|^2)w/(\kappa^2+w^2)$, $G_{w} = |g_{w}|^{2}\kappa/(\kappa^{2} +w^{2})$, $G_{u} = |g_{u}|^{2}\kappa/(\kappa^{2} +w^{2})$, and $G = G_w+G_u-\gamma$. 
All results presented hereafter are obtained within this adiabatic approximation.

In order to calculate the fluctuation operators at an arbitrary pulse duration time, we define annihilation operators of normalized temporal field modes~\cite{hofer2011quantum,he2014einstein}
\begin{eqnarray}
A_{w}^{\rm in} &= &e^{-i\phi}\sqrt{\frac{2G}{1-e^{-2G\tau}}}\int_{0}^{\tau}dt\, a_{w}^{\rm in}(t)e^{-(G-i\delta)t}, \nonumber\\ 
A_{u}^{\rm in} &=& e^{i\phi}\sqrt{\frac{2G}{1-e^{-2G\tau}}}\int_{0}^{\tau}dt\, a_{u}^{\rm in}(t)e^{-(G-i\delta)t} ,\nonumber\\
A_{w}^{\rm out} &=& e^{i\phi}\sqrt{\frac{2G}{e^{2G\tau}-1}}\int_{0}^{\tau}dt\, a_{w}^{\rm out}(t)e^{(G+i\delta)t}, \nonumber\\
 A_{u}^{\rm out} &= &e^{-i\phi}\sqrt{\frac{2G}{e^{2G\tau}-1}}\int_{0}^{\tau}dt\, a_{u}^{\rm out}(t)e^{(G+i\delta)t} ,\nonumber\\
A^{\rm in}_m &=& m(0) ,\   A^{\rm out}_m= m(\tau) e^{-i\delta \tau} ,\label{TemporalModes}
\end{eqnarray}
where $\tau$ is the interaction time with the laser pulses. The corresponding output modes $a_{w}^{\rm out}$ and $a_{u}^{\rm out}$ are obtained through the standard cavity input-output relation $a_{j}^{\rm out}=a_{j}^{\rm in} + \sqrt{2\kappa}a_{j}$ ($j=w,u$). 
It is easily verified that the normalized temporal operators obey the canonical commutation relations $[A_j^{\rm in},A_j^{\rm in \dagger}]=1$ and $[A_j^{\rm out},A_j^{\rm out \dagger}]=1$ ($j=w,u,m$).

With the definition of the normalized temporal field modes, the populations of the output modes are easily derived as 
\begin{eqnarray}
\left\langle (A_{w}^{\rm out})^{\dag}A_{w}^{\rm out}\right\rangle &=& \frac{G_w}{G}\Upsilon(r) ,\nonumber\\
\left\langle (A_{u}^{\rm out})^{\dag}A_{u}^{\rm out}\right\rangle &=& \frac{G_u}{G}\Upsilon(r) ,\nonumber\\
\left\langle (A_{m}^{\rm out})^{\dag}A_{m}^{\rm out}\right\rangle &=&n_0+ \Gamma(r) ,\label{Populations_wuModes}
\end{eqnarray}
and the correlation functions between different modes take the following form
\begin{eqnarray}
\left\langle (A_{w}^{\rm out})^{\dag}A_{u}^{\rm out}\right\rangle e^{i(\phi_{g_{w}}-\phi_{g_{u}})} &=&\left\langle (A_{u}^{\rm out})^{\dag}A_{w}^{\rm out}\right\rangle e^{-i(\phi_{g_{w}}-\phi_{g_{u}})} \nonumber\\
 &=&\frac{\sqrt{G_wG_u}}{G}\Upsilon(r) ,\nonumber\\
\left\langle A_{m}^{\rm out}A_{w}^{\rm out} \right\rangle e^{-i\phi_{g_{w}}}&=& -i\sqrt{\frac{G_w}{G}} \Lambda(r) ,\nonumber\\
\left\langle A_{m}^{\rm out}A_{u}^{\rm out} \right\rangle e^{-i\phi_{g_{u}}} &=& -i\sqrt{\frac{G_u}{G}} \Lambda(r) ,\nonumber\\
\left\langle (A_{w}^{\rm out})^{\dag}(A_{u}^{\rm out})^{\dag}\right\rangle &=& \left\langle A_{w}^{\rm out}A_{u}^{\rm out}\right\rangle = 0,\nonumber\\
\left\langle (A_{m}^{\rm out})^{\dag}A_{w}^{\rm out} \right\rangle &=&\left \langle (A_{m}^{\rm out})^{\dag}A_{u}^{\rm out} \right\rangle = 0 , 
\label{Correlations_wuModes}
\end{eqnarray}
where the coefficients denote
\begin{eqnarray}
\Gamma(r) &=& \left(e^{2r}-1\right)\left[(n_0+1)+\frac{\gamma}{G}(n_m+1)\right] ,\nonumber\\
\Upsilon(r) &=& 2\frac{\gamma}{G} (n_{m}+1)\frac{e^{2r} (\sinh2r - 2r)}{e^{2r}-1} \nonumber\\
&&+\left(n_{0}+1\right)\left(e^{2r}-1\right), \nonumber\\
\Lambda(r) &=& e^r\sqrt{e^{2r}-1}\Bigg[\frac{\gamma}{G}\left(n_m+1\right)\left(1-\frac{2r}{e^{2r}-1}\right) \nonumber\\
&&+n_0+1\Bigg] .
\label{Coefficients}
\end{eqnarray}
Here, we have assumed that the mechanical mode is initially in a thermal state with the mean phonon number $n_0$, i.e., $\langle m^\dagger(0)m(0)\rangle=n_0$. $r=G\tau$ represents an effective squeezing parameter, while $\phi_{g_{w}}$ and $\phi_{g_{u}}$ respectively denote the phases associated with the coupling strengths of the superposition modes, $g_{w}=|g_{w}|\exp(i\phi_{g_{w}})$ and $ g_{u}=|g_{u}|\exp(i\phi_{g_{u}})$.

We now proceed to analyze the populations and correlations of the orginal optical modes $a_1,~a_2$ as well as the mechanical mode $m$, and further construct the corresponding covariance matrix.
We begin by inverting the orthogonal transformation of  Eq.~(\ref{OrthogonalSuperpositionModes}), yielding
\begin{eqnarray}
a_{1} &=& a_{w}\cos\theta + a_{u}\sin\theta  , \nonumber \\
a_{2} &=& a_{w}\sin\theta - a_{u}\cos\theta .\label{OrthogonalSuperpositionModes_12}
\end{eqnarray}
Next, we define the annihilation operators of the normalized temporal output modes 
\begin{eqnarray}
A_{j}^{\rm out} = \sqrt{\frac{2G}{e^{2G\tau}-1}}\int_{0}^{\tau}dt\, a_{j}^{\rm out}(t)e^{(G+i\delta)t} ,\ j=1,2 .
\label{TemporalModes_12}
\end{eqnarray}
From Eqs.~(\ref{TemporalModes},~\ref{OrthogonalSuperpositionModes_12},~\ref{TemporalModes_12}), we
obtain the relationship between $A_{1}^{\rm out}$, $A_{2}^{\rm out}$ and $A_{w}^{\rm out}$, $A_{u}^{\rm out}$
\begin{eqnarray}
A_{1}^{\rm out} &=& A_{w}^{\rm out}e^{-i\phi}\cos\theta + A_{u}^{\rm out}e^{i\phi}\sin\theta ,\nonumber\\
A_{2}^{\rm out} &=& A_{w}^{\rm out}e^{-i\phi}\sin\theta - A_{u}^{\rm out}e^{i\phi}\cos\theta .
\label{TemporalModes_12_wu}
\end{eqnarray}
Therefore, the populations of the modes and the corresponding correlation functions are found to be
\begin{eqnarray}
	\left\langle (A_j^{\rm out})^{\dag}A_j^{\rm out}\right\rangle =  \frac{\kappa \Upsilon(r)}{G(\kappa^{2} +w^{2})}|{\cal A}_{j}(\psi)|^{2} , \quad  j=1,2, \nonumber\\
	\left\langle (A_1^{\rm out})^{\dag}A_2^{\rm out}\right\rangle e^{i(\phi_{{\cal A}_{1}}-\phi_{{\cal A}_{2}})} =\frac{\kappa \Upsilon(r)}{G(\kappa^{2} +w^{2})}|{\cal A}_{1}(\psi)||{\cal A}_{2}(\psi)| , \nonumber\\
	\left\langle A_m^{\rm out}A_j^{\rm out}\right\rangle e^{-i\phi_{{\cal A}_{j}}} = -i\sqrt{\frac{\kappa}{G(\kappa^2+w^2)}} \Lambda(r)\, |{\cal A}_{j}(\psi)| ,\nonumber\\
	\left\langle (A_{1}^{\rm out})^{\dag}(A_{2}^{\rm out})^{\dag}\right\rangle = \left\langle A_{1}^{\rm out}A_{2}^{\rm out}\right\rangle = 0, \, 
      \left \langle (A_{m}^{\rm out})^{\dag}A_{j}^{\rm out} \right\rangle = 0 ,\nonumber\\
\label{Populations_Correlations_12} 
\end{eqnarray}
where coefficients ${\cal A}_{1}(\psi)=g_w e^{-i\phi} \cos{\theta}+g_u e^{i\phi} \sin{\theta}$ and ${\cal A}_{2}(\psi)=g_w e^{-i\phi} \sin{\theta}-g_u e^{i\phi} \cos{\theta}$. The phases $\phi_{{\cal A}_{j}}$ are extracted from the complex amplitude ${\cal A}_{j}(\psi) = |{\cal A}_{j}(\psi)|\exp(i\phi_{{\cal A}_{j}})$. Correspondingly, the amplitudes takes a clear form
\begin{eqnarray}
\left|{\cal A}_{j}(\psi)\right|^{2} &=& \left|g_{1}\right|^{2} +(-1)^j \left(|g_{1}|^{2} -|g_{2}|^{2}\right)\sin^{2}2\theta\sin^{2}\phi \nonumber\\
&&-(-1)^j \left|g_{1}||g_{2}\right|\sin4\theta\sin^{2}\phi\cos2\psi  , \nonumber\\
&&+(-1)^j\left |g_{1}||g_{2}\right|\sin2\theta\sin2\phi\sin2\psi.
\label{Coefficient_A}
\end{eqnarray}
Here, $2\psi=\phi_{g_{1}}-\phi_{g_{2}}$ represents the relative phase between the two coupling strengths $g_1$ and $g_2$. Note that this relative phase would induce distinct interference effects, enabling phase‑controlled quantum entanglement and Bell nonlocality.

To determine the covariance matrix, we define the following quadrature components for the output field of  the cavity modes and the mechanical mode
\begin{eqnarray}
X^{\rm out}_{j} &=& \frac{1}{\sqrt{2}}\!\left[A^{\rm out}_{j}e^{-i\phi_{{\cal A}_{j}}} +(A^{\rm out}_{j})^{\dag}e^{i\phi_{{\cal A}_{j}}}\!\right]  ,\nonumber\\
P^{\rm out}_{j} &=& \frac{1}{\sqrt{2}i}\!\left[A^{\rm out}_{j}e^{-i\phi_{{\cal A}_{j}}} -(A^{\rm out}_{j})^{\dag}e^{i\phi_{{\cal A}_{j}}}\!\right] ,\ j=1,2, \nonumber\\
X^{\rm out}_{m} &=& \frac{1}{\sqrt{2}}\left[A^{\rm out}_{m} +(A^{\rm out}_{m})^{\dag}\right] ,\nonumber \\
P^{\rm out}_{m} &=& \frac{1}{\sqrt{2}i}\left[A^{\rm out}_{m} -(A^{\rm out}_{m})^{\dag}\right].
\label{Quadrature_OutputModes}
\end{eqnarray}
Similarly, we define the following quadrature components for the input field modes
\begin{eqnarray}
X^{\rm in}_{j} &=& \frac{1}{\sqrt{2}}\left[A^{\rm in}_{j} +(A^{\rm in}_{j})^{\dag}\right] ,\nonumber\\
P^{\rm in}_{j} &=& \frac{1}{\sqrt{2}i}\left[A^{\rm in}_{j} -(A^{\rm in}_{j})^{\dag}\right] ,\ j=1,2,m.
\label{Quadrature_InputModes}
\end{eqnarray}
By employing Eqs.~(\ref{Populations_wuModes}) and~(\ref{Populations_Correlations_12}), the variances of the output quadratures and their correlations can be readily obtained
\begin{gather}
\Delta^{2}X^{\rm out}_{m} = \Delta^{2}P^{\rm out}_{m} = n_{0} +\frac{1}{2} +\Gamma(r) ,\nonumber\\
\Delta^{2}X_{j}^{\rm out} = \Delta^{2}P_{j}^{\rm out} = \frac{1}{2} +\Upsilon(r)|{\cal U}_{j}(\psi)|^{2} ,\nonumber\\
\left\langle X_{m}^{\rm out} P_{j}^{\rm out}\right\rangle = \left\langle P_{m}^{\rm out} X_{j}^{\rm out}\right\rangle = -\Lambda(r)|{\cal U}_{j}(\psi)| ,\nonumber \\
\left\langle X_{1}^{\rm out} X_{2}^{\rm out}\right\rangle =\left \langle P_{1}^{\rm out} P_{2}^{\rm out}\right\rangle  =\Upsilon(r)|{\cal U}_{1}(\psi){\cal U}_{2}(\psi)|,
\label{Fluctuations_Quadratures}
\end{gather}
where 
\begin{eqnarray}
|{\cal U}_{j}(\psi)| =
\sqrt{\frac{\kappa}{G(\kappa^2+w^2)}}|{\cal A}_{j}(\psi)| ,\quad {\rm for}\quad  j=1,2 .  
\label{Coefficient_U}
\end{eqnarray}

Given the above populations and correlation functions, we can easily construct the covariance matrix of our proposed closed-loop three-mode optomechanical system, which is defined as
\begin{equation}
V=\left(
\begin{array}{ccc}
V_{1}  & V_{1,2} & V_{1,m} \\
V_{1,2}^T & V_{2} &   V_{2,m} \\
V_{1,m}^T & V_{2,m}^T & V_m 
\end{array}
\right).
\label{Eq_CovarianceMatrix}
\end{equation}
Here, the block matrices are 
\begin{eqnarray}
V_j&=&\left(
\begin{array}{cc}
V(X_j^{\rm out}) & 0  \\
0 & V(P_j^{\rm out})
\end{array}
\right), \nonumber \\
V_{j,k}&=&\left(
\begin{array}{cc}
V(X_j^{\rm out},X_k^{\rm out}) & V(X_j^{\rm out},P_k^{\rm out})  \\
V(P_j^{\rm out},X_k^{\rm out}) & V(P_j^{\rm out},P_k^{\rm out})
\end{array}
\right) ,
\end{eqnarray}
with the elements being defined as $V(O_j^{\rm out})=\Delta^2 O_j^{\rm out}=\langle (O_j^{\rm out})^2\rangle-\langle O_j^{\rm out} \rangle^2$, $V(O_j^{\rm out},O_k^{\rm out})=\langle O_j^{\rm out}O_k^{\rm out} +O_k^{\rm out}O_j^{\rm out}\rangle/2-\langle O_j^{\rm out}\rangle\langle O_k^{\rm out}\rangle$ ($O=X,P$ and $j,k=1,2,m$). Hence, the covariance matrix can be straightforwardly constructed from Eq.~(\ref{Fluctuations_Quadratures}).

\section{The criteria of verifying entanglement and Bell nonlocality}
\label{Criteria}
In the previous section, we derived an analytical expression for the covariance matrix of the closed-loop optomechanical system from its Hamiltonian and equations of motion. We now present the criteria for verifying entanglement and Bell nonlocality in both bipartite and tripartite systems.

\subsection{The criteria of verifying bipartite and tripartite entanglement}
In order to measure the bipartite entanglement for arbitrary Gaussian states, we adopt the widely used logarithmic negativity~\cite{Vidal2002,Adesso2004}, which is defined as 
\begin{equation}\label{QuanMagnonics_LogNegativity}
E_N\equiv \max{\left\{0,-\ln{2\eta^{-}}\right\}},
\end{equation}
with $\eta^{-}= \min \mathrm{eig}\left | i\Omega_2 \widetilde{V}_{\mathrm{reduced}}  \right|$ being the minimum symplectic eigenvalue of the partial transposed covariance matrix $\widetilde{V}_{\mathrm{reduced}}=PV_{\mathrm{reduced}}P$. Here, $\Omega_2=\bigoplus_{i=1}^2 i\hat{\sigma}_y$ represents the symplectic matrix, where $\hat{\sigma}_y$ denotes the Pauli operator. $P=\mathrm{diag}(1,-1,1,1)$ is used for implementing partial transposition. 
And $V_{\mathrm{reduced}}$ denotes the $4\times4$ reduced covariance matrix of the concerned two-mode subsystem, obtained by removing the rows and columns of the uninteresting mode from the covariance matrix of the whole system. For example, in our proposal, since we are specifically interested in the bipartite entanglement created between the mechanical mode $m$ and the optical mode $a_j~(j=1,~2)$, it is enough to consider the reduced covariance matrix
\begin{equation}
V_{\mathrm{reduced}}^{m\oplus j}=\left(
   \begin{array}{cc}
   V_j & V_{j,m}\\
   V_{j,m}^T & V_m
   \end{array}
   \right), 
\end{equation}
where the definitions of three submatrices $V_j$, $V_m$ and $V_{j,m}$ are referred to Eq.~(\ref{Eq_CovarianceMatrix}). 
Therefore, in a two-mode Gaussian subsystem, bipartite entanglement is present if and only if $\eta^{-}<1/2$. The value of logarithmic negativity quantifies the degree of entanglement, i.e., a higher value of $E_N$ corresponds to a stronger degree of bipartite entanglement.

To signify genuine tripartite entanglement in the three-mode Gaussian continuous-variable system, the residual contangle $E_\tau$ is generally employed~\cite{Gerardo2006NJP_TripartiteEnt,Gerardo2007JPA_EntReview}, which is conceptually analogous to the tangle quantifying tripartite entanglement in discrete-variable systems. The residual contangle takes the form of 
\begin{equation}\label{Eq_QuanMagnonics_ResContangle}
E_\tau^{u \mid v w}=C_\tau^{u \mid v w}-C_\tau^{u \mid v}-C_\tau^{u \mid w} \; (u,~v,~w=j,~k,~l), 
\end{equation}
where the symbols $\{j,~k,~l\}$ respectively represent the three modes of the concerning tripartite state and superscripts $\{u,~v,~w\}$ represent all the possible results when permuting three subsystems. 
$C_\tau^{u \mid v w}=(E_N^{u \mid v w})^2,~C_\tau^{u \mid v}=(E_N^{u \mid v})^2$ and $C_\tau^{u \mid w}=(E_N^{u \mid w})^2$ denote the contangle of subsystems, which are a proper entanglement monotone defined by a squared logarithmic negativity. Here, the one-mode-versus-one-mode logarithmic negativity $E_N^{u \mid v}$ and $E_N^{u \mid w}$ can be directly calculated referring to Eq.~(\ref{QuanMagnonics_LogNegativity}). Whereas, the one-mode-versus-two-modes logarithmic negativity $E_N^{u \mid v w}$ can be determined following the definition of Eq.~(\ref{QuanMagnonics_LogNegativity}) by simply replacing the symplectic matrix and partial transposition matrices. Specifically, $E_N^{u \mid v w}\equiv\max{\{0,-\ln{2(\eta^{-})^{u \mid v w}}\}}$, where the minimum symplectic eigenvalue reads $(\eta^{-})^{u \mid v w} = \min \mathrm{eig}\left | i \Omega_3 P^{u \mid v w} V' P^{u \mid v w}  \right|$. $V'$ is the corresponding $6\times6$ covariance matrix of the tripartite system. The symplectic matrix is redefined as $\Omega_3=\bigoplus_{i=1}^3 i\hat{\sigma}_y$. And the partial transposition matrices respectively denote
$P^{j \mid k l} =\mathrm{diag}(1,-1,1,1,1,1)$, $P^{k \mid j l} =\mathrm{diag}(1,1,1,-1,1,1)$ and $P^{l \mid j k} =\mathrm{diag}(1,1,1,1,1,-1)$.

It is easily found that the residual contangle defined in Eq.~(\ref{Eq_QuanMagnonics_ResContangle}) is not generally invariant under all permutations of the modes. 
Therefore, a \textit{bona fide} quantification for genuine tripartite entanglement of a three-mode Gaussian continuous-variable system is then given by the \textit{minimum} residual contangle~\cite{Gerardo2006NJP_TripartiteEnt,Gerardo2007JPA_EntReview}
\begin{equation}\label{Eq_QuanMagnonics_MinResContangle}
E_\tau^{\min } \equiv \min \left[E_\tau^{j \mid k l}, E_\tau^{k \mid j l}, E_\tau^{l \mid j k}\right].
\end{equation}
The above definition ensures that tripartite entanglement is invariant under all permutations of the modes, therefore, is a genuine three-way property of arbitrary three-mode Gaussian state.

\subsection{The criteria of verifying bipartite and tripartite Bell nonlocality}

In continuous-variable systems, Bell nonlocality can be detected through displaced-parity measurement, which is related to the Wigner function of the system ~\cite{LiJie_2017PRA_OptoBell,Banaszek1998PRA_Nonlocality,Jeong2003PRA_MultipartiteBell,vanLoock_2001PRA_GHZBell}.
For an $n$-mode continuous-variable Gaussian state with zero mean values, the Wigner function is fully determined by the covariance matrix $V$~\cite{LiJie_2017PRA_OptoBell,Gerardo_2014_CVSystem}:
\begin{equation}
W(u)=\frac{e^{-uV^{-1}u^T}}{\pi^n\sqrt{\det{V}}},
\label{WignerFunction}
\end{equation}
where $n$ denotes the number of modes in the concerned multipartite system, and $u= \{u_1, u_2, \dots, u_n\}$ represents the orthogonal fluctuation operators in phase space with $u_j= \{\delta  q_j, \delta  p_j\}$ ($j=1,~2,~\dots,~n$).

The displaced-parity operator for the $j$th mode $b_j$ is defined as $\hat{\Pi}_j(\alpha_j)=\hat{D}_j(\alpha_j)\hat{\Pi}_j\hat{D}_j^\dagger(\alpha_j)$, where $\hat{D}_j(\alpha_j)=\exp\left(\alpha_j b_j^\dagger-\alpha_j^*b_j\right)$ denotes the displacement operator, and $\hat{\Pi}_j=(-1)^{\hat{n}_j}=\sum_{n=0}^{\infty}(|2n\rangle\langle2n|-|2n+1\rangle\langle2n+1|)$ represents the parity operator with $\hat{n}_j=b_j^\dagger b_j$. Thus, the displaced-parity operator for the $n$-mode system takes the form of $\hat{\Pi}(\alpha)=\bigotimes_{j=1}^{n}\hat{\Pi}_j(\alpha_j)$, where $\alpha=(\alpha_1, \alpha_2, \dots, \alpha_n)$ denotes the vector of displacement amplitudes for the respective modes. Experimentally, the displaced‑parity measurement can be implemented by combining a beam splitter with a photon‑number‑resolving detector. 
Specifically, the signal mode is interfered with a coherent reference field (the displacement), and the output ports are monitored for photon‑number parity~\cite{Banaszek1996PRL}. This setup effectively maps the parity of the displaced field onto the probabilities of detecting even and odd photon numbers, which can be repeated for different displacement amplitudes to sample the Wigner function point‑by‑point. 
It is important to note that a direct displaced‑parity measurement on the mechanical resonators is experimentally challenging. Instead, the mechanical state can be transferred to an additional optical mode via a weak red‑detuned probe light, after which the displaced‑parity measurement is performed on the ancillary optical mode~\cite{Vitali2007PRL_OptoEnt,Palomaki710}. This indirect readout scheme leverages well‑developed optical detection techniques while preserving the relevant quantum correlations.

The displaced-parity operator connects directly to the Wigner function via the following identity~\cite{LiJie_2017PRA_OptoBell,Banaszek1998PRA_Nonlocality}
\begin{equation}
\left<\hat{\Pi}(\alpha)\right>=\frac{\pi}{2}W(\alpha).
\label{DisplacedParity_Wigner_Relationship}
\end{equation}
Such relation forms the basis for constructing Bell inequalities in phase space. The phase space version of Bell-Clauser-Horne-Shimony-Holt (CHSH) inequality for a bipartite subsystem~\cite{Clauser1969Proposed} can be expressed as
\begin{eqnarray}
\mathcal{B}_2&=&\frac{\pi^2}{4}\bigg[W(u_1,u_2)+W(u_1',u_2)+W(u_1,u_2') \nonumber \\
&&-W(u_1',u_2')\bigg].
\end{eqnarray}
Corresponding, Mermin-Klyshko-type Bell inequality for a tripartite system~\cite{Mermin1990Extreme,Klyshko1993Bell} takes the form of 
\begin{eqnarray}
\mathcal{B}_3&=&\frac{\pi^3}{8}\bigg[W(u_1,u_2,u_3')+W(u_1,u_2',u_3)+W(u_1',u_2,u_3)\nonumber\\
&&-W(u_1',u_2',u_3')\bigg].
\end{eqnarray}
Here, $u_j$ and $u_j'$ ($j=1,~2,~3$) denote two distinct phase‑space settings for each party. In practice, the optimal settings are aligned with the squeezed and anti‑squeezed quadratures determined by the covariance matrix $V$ to maximize the correlation contrast.

Any local realistic theory imposes the bound $|\mathcal{B}_2|\leq 2$ and $|\mathcal{B}_3|\leq 2$ for bipartite and tripartite systems, respectively. Therefore, the violation of the Bell CHSH inequality, i.e., $|\mathcal{B}_2|>2$ and $|\mathcal{B}_3|>2$, respectively indicates the presence of Bell nonlocality shared between the relevant bipartite and tripartite systems. The maximal violation permitted by quantum mechanics are respectively $|\mathcal{B}_2|_{\max}=2\sqrt{2}$ and $|\mathcal{B}_3|_{\max}=4$~\cite{Cirelson_1980LettMathPhys_Bell,Gisin_1998_Bell}.

\begin{figure*}[htbp!]
\centering
\includegraphics[width=2.0\columnwidth]{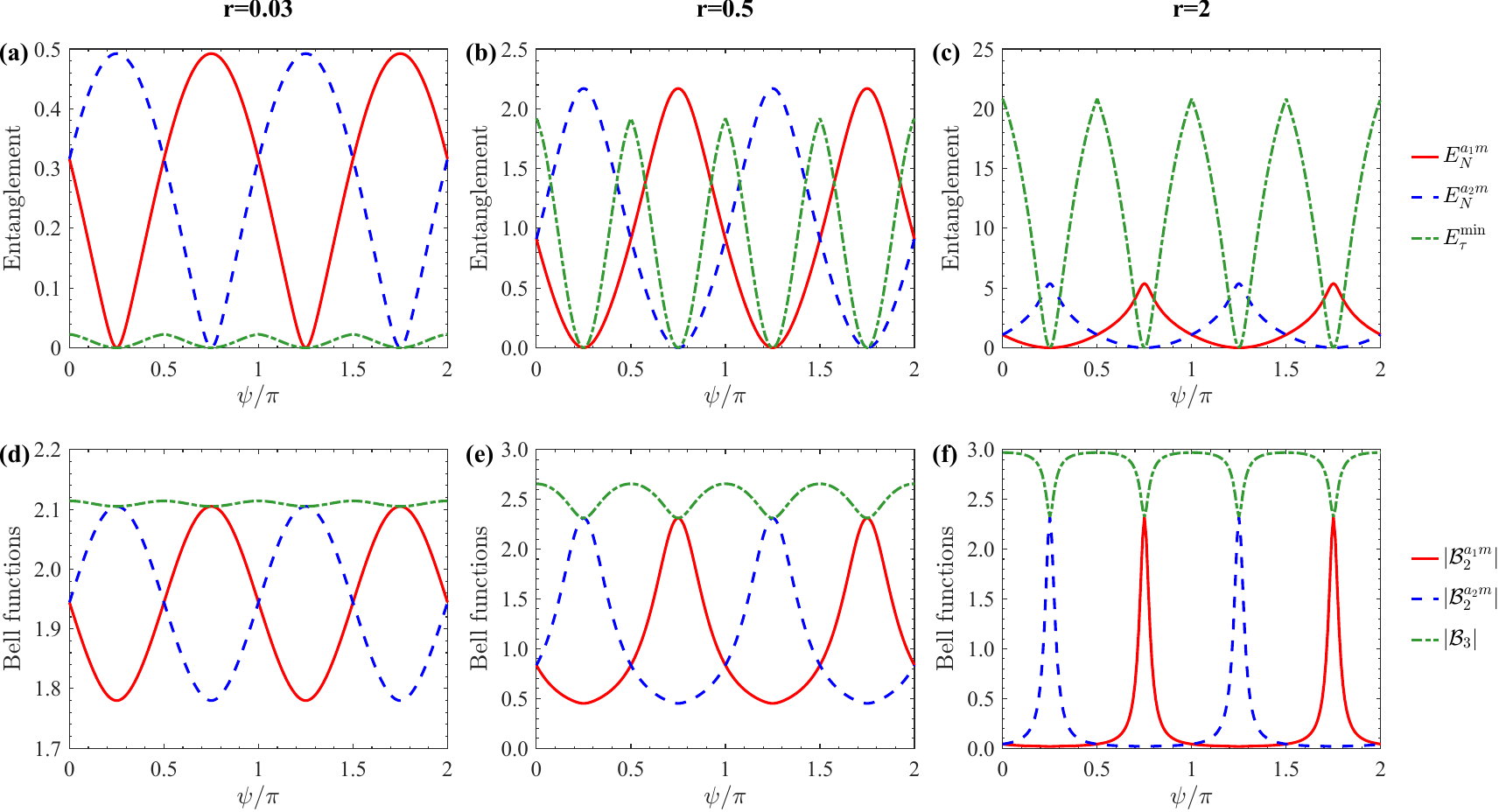}
\caption{(a-c) Bipartite and genuine tripartite entanglement and (d-f) the corresponding Bell functions versus the relative phase $\psi$ of the closed-loop three-mode optomechanical system. Specifically, $E_N^{a_j m}$($j=1,~2$) represents the bipartite entanglement between the cavity mode $a_j$ and the mechanical mode $m$, while $E_\tau^{\min }$ quantifies the genuine tripartite entanlement among all three modes. Besides, $\left|\mathcal{B}_2^{a_j m}\right|$ is the maximized bipartite Bell function of the $a_j-m$ subsystem under displaced partiy measurement through optimizing the quadratures $\left\{ u_j,~u_j' \right\}$($j=1,~2$) , and $\left|\mathcal{B}_3\right|$ indicates the corresponding maximized tripartite Bell function of the tripartite system by optimizing the quadratures $\left\{ u_j,~u_j' \right\}$($j=1,~2,~3$).
Each column corresponds to a fixed squeezing parameter $r$, with values $r = 0.03$, $0.5$, and $2$ from left to right. 
The remaining parameters denote $g = 0.1 \times 10^{6}$Hz, $\gamma = 0$, $\kappa = 10 g$, $n_0 = 0$, $n_m = 0$, $g_1 = g_2 = g$, $J = 10 g$, $\Delta = 0$. It should be emphasized that the parameters considered here are experimentally feasible~\cite{purdy2013observation,sankey2010strong,heinrich2011dynamics,fink2016quantum,barzanjeh2017mechanical,peterson2016laser}.}
\label{Fig_FunctionPsi}
\end{figure*}

\section{Numerical Results}\label{NumericalResults}
In the previous section, we established the theoretical framework of the proposed closed-loop three-mode optomechanical system and further recalled the criteria of verifying entanglement and Bell nonlocality. In this section, we will thoroughly investigate the properties of the bipartite and tripartite entanglement and Bell nonlocality.

\subsection{Phase control of bipartite entanglement and Bell nonlocality}
\label{Results_PhaseControlBipartiteEntBell}

In this subsection, we first discuss the effects of the relative phase $\psi$ on both bipartite entanglement and Bell nonlocality. 
As demonstrated in the Hamiltonian~(\ref{Hamiltonian}), two optical modes $a_1$ and $a_2$ are coupled through a beam‑splitter‑type interaction, preventing any generation of bipartite entanglement and Bell nonlocality between them. We therefore focus exclusively on the bipartite quantum correlations between each optical mode with the mechanical mode, as well as tripartite quantum correlations among three modes within the system. 
Figure~\ref{Fig_FunctionPsi} displays the bipartite and tripartite entanglement along with Bell functions against the relative phase $\psi$ for different effective squeezing parameter $r$, in which $E_N^{a_j m}$ and $|\mathcal{B}_2^{a_j m}|$ ($j=1,~2$) respectively represent the bipartite entanglement and the maximized bipartite Bell functions between the optical mode $a_j$ and the mechanical mode $m$. Besides, $E_\tau^{\mathrm{min}}$ and $|\mathcal{B}_3|$ respectively denote the genuine tripartite entanglement and the maximized tripartite Bell functions for the whole tripartite system. Note that both bipartite and tripartite Bell functions displayed in this manuscript are optimized to the maximal value by optimizing the corresponding quadratures $\left\{ u_j,~u_j' \right\}$ ($j=1,~2$ for bipartite subsystems and $j=1,~2,~3$ for tripartite system).
As is evident from Fig.~\ref{Fig_FunctionPsi}, all quantities oscillate periodically with the relative phase $\psi$. Specifically, the bipartite entanglement and Bell functions both exhibit a period of $\pi$, whereas the corresponding tripartite measures vary with a period of $\pi/2$. This phase-controlled behavior is also independent of the squeezing parameter $r$, as evidenced by the consistent patterns across the three columns.
Furthermore, it can be apparently observed from Figs.~\ref{Fig_FunctionPsi}(a-c) that the bipartite entanglement $E_N^{a_1 m}$ (shown by red solid curves) and $E_N^{a_2 m}$ (shown by blue dashed curves) are competitive. The bipartite entanglement $E_N^{a_1 m}$ and $E_N^{a_2 m}$ can be periodically transferred between the modes through manipulating the relative phase $\psi$.
To be specific, when the relative phase $\psi = (n + 3/4)\pi$ ($n = 0, 1, 2,\dots$), the bipartite entanglement $E_N^{a_1 m}$ between the cavity mode $a_1$ and the mechanical mode $m$ reaches its maximum value, while there is no entanglement between the cavity mode $a_2$ and the mechanical mode $m$ (i.e., $E_N^{a_2 m} = 0$). 
Corresponding, when the relative phase $\psi = (n + 1/4)\pi$ ($n = 0, 1, 2,\dots$), the bipartite entanglement $E_N^{a_2 m}$ between the cavity mode $a_2$ and the mechanical mode $m$ reaches its maximum value, while no entanglement exists between the cavity mode $a_1$ and the mechanical mode $m$ (i.e., $E_N^{a_1 m} = 0$). 
When the relative phase $\psi = n\pi/2$ ($n = 0, 1, 2,\dots$), the bipartite entanglement $E_N^{a_1 m}$ is exactly equal to $E_N^{a_2 m}$. 

The underlying physical mechanism behind this phase‑controlled behavior can be understood as follows. For the chosen parameters listed in the caption of Fig.~\ref{Fig_FunctionPsi}, we have $\cos^{2}\theta = 1/2$ and $\tan\phi = 1$, so the coefficient defined in Eq.~\eqref{Populations_Correlations_12} reduces to
\begin{equation}
|\mathcal{A}_j(\psi) |^2= g^{2} \left[1 + \left(-1\right)^{j} \sin 2\psi\right ], \quad (j=1,2).
\label{Populations_Correlations_12_Simplified}
\end{equation}
When the relative phase $\psi = (n+3/4)\pi$ ($n=0,1,2,\dots$), the coefficients simplify to $|\mathcal{A}_1(\psi)|=\sqrt{2}g$ and $|\mathcal{A}_2(\psi)|=0$, with $|\mathcal{A}_1(\psi)|$ reaching its maximum. Consequently,  the output mode $a_2$ is completely depopulated, $\langle (A_2^{\mathrm{out}})^{\dagger}A_2^{\mathrm{out}}\rangle = 0$, whereas the output mode $a_1$ achieves maximum population. This population imbalance yields maximal bipartite entanglement between $a_1$ and $m$, and no entanglement between $a_2$ and $m$. 
Conversely, when the phase $\psi = (n+1/4)\pi$ ($n=0,1,2,\dots$), the coefficients are respectively $|\mathcal{A}_1(\psi)|=0$ and $|\mathcal{A}_2(\psi)|=\sqrt{2}g$, where $|\mathcal{A}_2(\psi)|$ reaches its maximum. Then the situation is totally reversed, that is, $a_1$ is unpopulated and $a_2$ is maximally populated, leading to maximal $a_2\!-\!m$ entanglement and vanishing $a_1\!-\!m$ entanglement.
At the symmetric points $\psi = n\pi/2$ ($n=0,1,2,\dots$), the two coefficients coincide, $|\mathcal{A}_1(\psi)| =| \mathcal{A}_2(\psi)|$, resulting in equal populations of two optical modes and hence identical bipartite entanglement $E_N^{a_1 m} = E_N^{a_2 m}$.
The above results clearly demonstrate that the relative phase $\psi$ directly governs the population distribution between the two optical modes, thereby determining the bipartite entanglement distribution among the optical-mechanical pairs. This phase-controlled entanglement redistribution  is in complete agreement with previous findings for loop‑coupled optomechanical systems~\cite{Sun_2017NJP}.

Comparing Figs.~\ref{Fig_FunctionPsi}(a-c) with different effective squeezing parameters $r$, we observe that the peak values of bipartite entanglement $E_N^{a_1 m}$ and $E_N^{a_2 m}$ grow monotonically with increasing $r$.
For instance, at a weak squeezing parameter $r=0.03$, the maximum attainable bipartite entanglement remains small at $E_N\approx 0.49$. In contrast, when the squeezing parameter is increased to  $r=2$, the maximum bipartite entanglement rises to $E_N\approx 5.38$, revealing a strong enhancement of quantum correlations induced by the effective squeezing parameter. 
This observation is substantiated by the analytical expressions in Eqs.~(\ref{Appendix_TriPartiteState_OneToOneEnt_Psi0d75}) and~(\ref{Appendix_TriPartiteState_OneToOneEnt_Psi0d25}) of Appendix~\ref{AppendixA}.
Specifically, at the optimal phases $\psi = (n+1/4)\pi$ and $\psi = (n+3/4)\pi$ ($n=0,1,2,\dots$), the peak values of bipartite entanglement are given by $E_N^{a_2 m} = 2\ln\left(e^{r}+\sqrt{e^{2r}-1}\right)$ and $E_N^{a_1 m} = 2\ln\left(e^{r}+\sqrt{e^{2r}-1}\right)$, respectively, which indeed increase monotonically with $r$.
However, this monotonic dependence on $r$ only holds under ideal conditions without mechanical dissipation and thermal noises (i.e., $\gamma=0,~n_m=0,~n_0=0$). When mechanical dissipation and thermal noises are present,
the influence of squeezing parameter $r$ becomes more intricate, which will be systematically investigated in Sec.~\ref{Results_Influence_DissipationLoss}.

Having observed strong bipartite entanglement, we further investigate the bipartite Bell nonlocality for the $a_1\!-\!m$ and $a_2\!-\!m$ subsystems. The bipartite Bell functions $|\mathcal{B}_2^{a_1 m}|$ (shown by red solid curves) and $|\mathcal{B}_2^{a_2 m}|$ (shown by blue dashed curves) exhibit phase dependence that closely follow those of the corresponding bipartite entanglement $E_N^{a_1 m}$ and $E_N^{a_2 m}$, respectively. 
Specifically, $|\mathcal{B}_2^{a_1 m}|$ attains its maximum when $\psi = (n+3/4)\pi$ ($n=0,1,2,\dots$), coinciding precisely with the phase that maximizes the bipartite entanglement $E_N^{a_1 m}$. Similarly, $|\mathcal{B}_2^{a_2 m}|$ reaches its peak at $\psi = (n+1/4)\pi$ ($n=0,1,2,\dots$), the same phase that optimizes bipartite entanglement $E_N^{a_2 m}$. This one‑to‑one correspondence between the phases maximizing bipartite entanglement and those maximizing the bipartite Bell functions underscores the intrinsic connection between these two non‑classical signatures in the considered loop‑coupled optomechanical system.
Under the chosen parameters listed in the caption of Fig.~\ref{Fig_FunctionPsi}, however, the bipartite Bell inequalities cannot be violated for all values of $\psi$. Violations ($|\mathcal{B}_2^{a_1 m}|>2,~|\mathcal{B}_2^{a_2 m}|>2$) occur only within limited phase windows centered around the corresponding optimal points mentioned above. 
For a small squeezing parameter $r=0.03$, as manifested by the red solid and blue dashed curves in Fig.~\ref{Fig_FunctionPsi}(d),  the violation windows are relatively broad, spanning approximately  $\Delta \psi_1=  (0.56 + n)\pi \sim (0.94 + n)\pi$ for $|\mathcal{B}_2^{a_1 m}|$ and $\Delta \psi_2= (0.06 + n)\pi \sim (0.44 + n)\pi$ for $|\mathcal{B}_2^{a_2 m}|$ ($n = 0,1,2,\dots$).
However, the maximal bipartite Bell violations of both $|\mathcal{B}_2^{a_1 m}|$ and $|\mathcal{B}_2^{a_2 m}|$ reach only $|\mathcal{B}_2|_{\max}\approx2.10$. In contrast, when the effective squeezing parameter increases to $r = 2$, as displayed by the red solid and blue dashed curves in Fig.~\ref{Fig_FunctionPsi}(f), the violation windows for the bipartite Bell functions $|\mathcal{B}_2^{a_1 m}|$ and $|\mathcal{B}_2^{a_2 m}|$ are markedly narrower than that for smaller squeezing parameters, which are respectively 
$\Delta \psi_1'= (0.738+ n)\pi \sim (0.762+ n)\pi$ and $\Delta \psi_2'= (0.238+ n)\pi \sim (0.262+ n)\pi$ ($n = 0, 1, 2,\dots$), while the peak violation rises to $|\mathcal{B}_2|_{\max}\approx2.32$. 
These results reveal a clear trade‑off, that is, increasing the effective squeezing parameter $r$ enhances the maximal violation strength but narrows the phase interval over which violation occurs, while decreasing $r$ broadens the violation window at the expense of a weaker violation strength. This tunable compromise between the magnitude of Bell-inequality violation and the range of the phase parameter provides a practical handle for tailoring bipartite Bell nonlocality to specific experimental requirements.

\subsection{Phase control of tripartite entanglement and Bell nonlocality}
\label{Results_PhaseControlTripartiteEntBell}
Following the analysis of phase control effects on bipartite entanglement and Bell nonlocality in the previous subsection, we now examine how the relative phase $\psi$ influences the tripartite correlations in the closed‑loop three-mode optomechanical system.

In contrast to the bipartite correlations, which oscillate with period $\pi$, the tripartite entanglement and Bell functions vary with period $\pi/2$, as manifested by the green dash-dotted curves in Fig.~\ref{Fig_FunctionPsi}.
It is apparently observed from Figs.~\ref{Fig_FunctionPsi}(a-c) that the minimum residual contangle $E_\tau^{\mathrm{min}}$ quantifying genuine tripartite entanglement vanishes at phases $\psi = (n/2 + 1/4)\pi$ ($n=0,1,2,\dots$). This behavior is analytically confirmed by Eq.~(\ref{Appendix_TriPartiteState_MinimalResidualContangle_Psi0d75}) in Appendix~\ref{AppendixA_Psi0d75}. 
The origin of this phenomenon lies in the distribution of bipartite entanglement. At these specific phases, one of the bipartite entanglement $E_N^{a_j m}$ ($j=1,2$) reaches its maximum while the other vanishes, indicating that the mechanical mode is entangled exclusively with a single optical mode. Such an unbalanced distribution of bipartite correlations precludes genuine tripartite entanglement, yielding $E_\tau^{\mathrm{min}} = 0$.
Conversely, $E_\tau^{\mathrm{min}}$ attains its maximal values at the symmetric phases $\psi = n\pi/2$ ($n=0,1,2,\dots$), with the analytical expression derived in detail in Eq.~(\ref{Appendix_TriPartiteState_MinimalResidualContangle_Psi0}) of Appendix~\ref{AppendixA_Psi0}. According to the analysis following Eq.~(\ref{Populations_Correlations_12_Simplified}) in the preceding section, these symmetric phases correspond to equal populations of the two optical modes, which yields identical bipartite entanglement $E_N^{a_1 m} = E_N^{a_2 m}$. This balanced distribution of pairwise correlations leads to the optimal enhancement of genuine tripartite entanglement among all three modes. 
The above results demonstrate that the relative phase $\psi$ provides an effective method of precisely manipulating the degree of genuine tripartite entanglement in the proposed closed‑loop system. Such deterministic phase‑tunable generation and manipulation of multipartite entanglement is directly relevant to quantum‑information applications, including reconfigurable quantum networks~\cite{Wehner2018Science_QuantumInternet} and adaptive multi‑party protocols~\cite{Epping2017NJP_MultiParty} that require on‑demand entanglement distribution.

As shown in Figs.~\ref{Fig_FunctionPsi}(d-f), the tripartite Bell function $|\mathcal{B}_3|$ exhibits the same periodic dependence on the relative phase $\psi$ as the genuine tripartite entanglement $E_\tau^{\mathrm{min}}$.  In particular, $|\mathcal{B}_3|$ attains its minima at phases $\psi = (n/2 + 1/4)\pi$ ($n=0,1,2,\dots$), while reaches its maxima at the symmetric cases when $\psi = n\pi/2$ ($n=0,1,2,\dots$).
Whereas the bipartite Bell violation occurs only within a limited phase interval that narrows as the squeezing parameter $r$ increases, the tripartite Bell function $|\mathcal{B}_3|$ exhibits violation across the full range of the phase $\psi$ under the chosen parameters. Nonetheless, the peak value of $|\mathcal{B}_3|$ at the optimal phase shows a similar dependence on $r$, growing monotonically with the squeezing parameter. For instance, as indicated in Fig.~\ref{Fig_FunctionPsi}(d), the maximal violation of tripartite Bell function reaches $|\mathcal{B}_3| \approx 2.11$ when the effective squeezing parameter $r=0.03$. And the tripartite Bell function rises to $|\mathcal{B}_3|\approx 2.97$ when the squeezing parameter increases to $r=2$, as displayed in Fig.~\ref{Fig_FunctionPsi}(f).

A notable feature emerges from the comparison between the genuine tripartite entanglement and the tripartite Bell nonlocality in Fig.~\ref{Fig_FunctionPsi}. It is apparently observed that at phases $\psi = (n/2 + 1/4)\pi$ ($n=0,1,2,\dots$), where only one optical mode is entangled with the mechanical mode and the genuine tripartite entanglement is absent, the tripartite Bell function $|\mathcal{B}_3|$ nevertheless exhibits  a significant violation. Moreover, the tripartite Bell function coincides exactly with one of the bipartite Bell functions at these phases, which is detailed theoretically confirmed in Appendix~\ref{AppendixA_Psi0d75}.
Specifically, when $\psi = (n + 1/4)\pi$ ($n=0,1,2,\dots$), where the bipartite entanglement exists solely in the $a_2\!-\!m$ subsystem, we have $|\mathcal{B}_3| = |\mathcal{B}_2^{a_2 m}|$, as shown by Eq.~(\ref{Appendix_TriPartiteState_BellFuncBiTriOpt_Psi0d75}). Conversely, when $\psi = (n + 3/4)\pi$ ($n=0,1,2,\dots$), where the bipartite entanglement resides only in the $a_1\!-\!m$ subsystem, the equality $|\mathcal{B}_3| = |\mathcal{B}_2^{a_1 m}|$ holds, as indicated by Eq.~(\ref{Appendix_TriPartiteState_BellFuncBiTriOpt_Psi0d25}).
Based on both the numerical results and the theoretical derivation presented above, we therefore conclude that the violation of the tripartite Mermin-Klyshko-type Bell inequality does not require the genuine tripartite entanglement.

To gain deeper insight into the relationship between the tripartite Bell nonlocality and genuine tripartite entanglement, we consider a simple three‑mode system where modes $A$ and $B$ are prepared in a two‑mode squeezed vacuum state $|r\rangle = \exp\{r(A^\dagger B^\dagger - AB)\}|0\rangle_A \otimes |0\rangle_B$ (squeezing factor $r \in \mathbb{R}$), while the mode $C$ remains separable from both modes $A$ and $B$. Appendix~\ref{Appendix_ArtificalThreeModes} contains a full description of this three-mode system and provides complete derivations of the bipartite and tripartite entanglement as well as the corresponding Bell functions. 
According to the analytical results presented in Eqs.~(\ref{Artifical_TriPartiteState_OneToOneEnt}) and (\ref{Artifical_TriPartiteState_MinimalResidualContangle}) of Appendix~\ref{Appendix_ArtificalThreeModes_Ent}, this artificially prepared state exhibits solely bipartite entanglement between modes $A$ and $B$, quantified by $E_N^{12}=2r$, while the genuine tripartite entanglement vanishes, i.e., $E_{\tau}^{123}=0$.  Numerical simulations for both bipartite and tripartite entanglement are presented in Fig.~\ref{Fig_ArtificialThreeMode}(a), which confirm the above analytical predictions with excellent agreement.
Nevertheless, the analytical result of Eq.~(\ref{Artifical_TriPartiteState_BiTriBell}) in Appendix~\ref{Appendix_ArtificalThreeModes_Bell} show that the tripartite Bell function $|\mathcal{B}_{123}|$ coincides exactly with the bipartite function $|\mathcal{B}_{12}|$ for all values of $r$, as visually confirmed in Fig.~\ref{Fig_ArtificialThreeMode}(c). Under ideal conditions, both bipartite and tripartite Bell functions grow monotonically with $r$ and exhibit violation whenever $r > 0$  even though the genuine tripartite entanglement is absent.
This constructed three‑mode example therefore definitively substantiates that the genuine tripartite entanglement is not a prerequisite for tripartite Bell nonlocality, demonstrating that tripartite Bell inequalities capture correlations beyond genuine multipartite entanglement.

\subsection{Effects of the squeezing parameter $r$ on entanglement and Bell nonlocality}
\label{Results_Influence_SqueezingParameter}

\begin{figure}
\centering
\includegraphics[width=0.96\columnwidth]{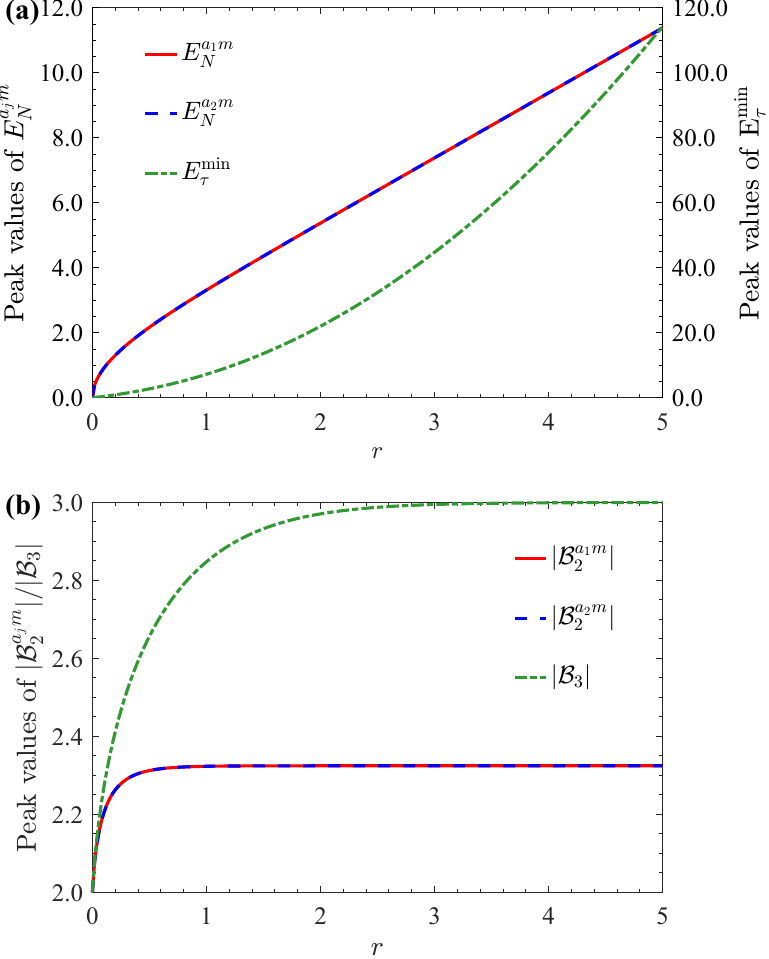}
\caption{(a) Maximized bipartite/tripartite entanglement and (b) bipartite/tripartite Bell functions plotted against the effective squeezing parameter $r$. All quantities are optimized to their maximal values by selecting the appropriate relative phase $\psi$. Specifically, the bipartite quantities $E_N^{a_1 m}$ and $\left|\mathcal{B}_2^{a_1 m}\right|$ are maximized at phases $\psi = (n + 3/4)\pi$ ($n=0,1,2,\dots$), while $E_N^{a_2 m}$ and $\left|\mathcal{B}_2^{a_2 m}\right|$ achieve their optimal values when $\psi = (n+ 1/4)\pi $ ($n=0,1,2,\dots$). The tripartite measures $E_\tau^{\mathrm{min}}$ and $|\mathcal{B}_3|$ are optimized at $\psi = n\pi/2$ ($n=0,1,2,\dots$). Unless otherwise stated, all other parameters remain the same as those in Fig.~\ref{Fig_FunctionPsi}.}
\label{Fig_FunctionR_NoLoss}
\end{figure}
\begin{figure*}
\centering
\includegraphics[width=1.5\columnwidth]{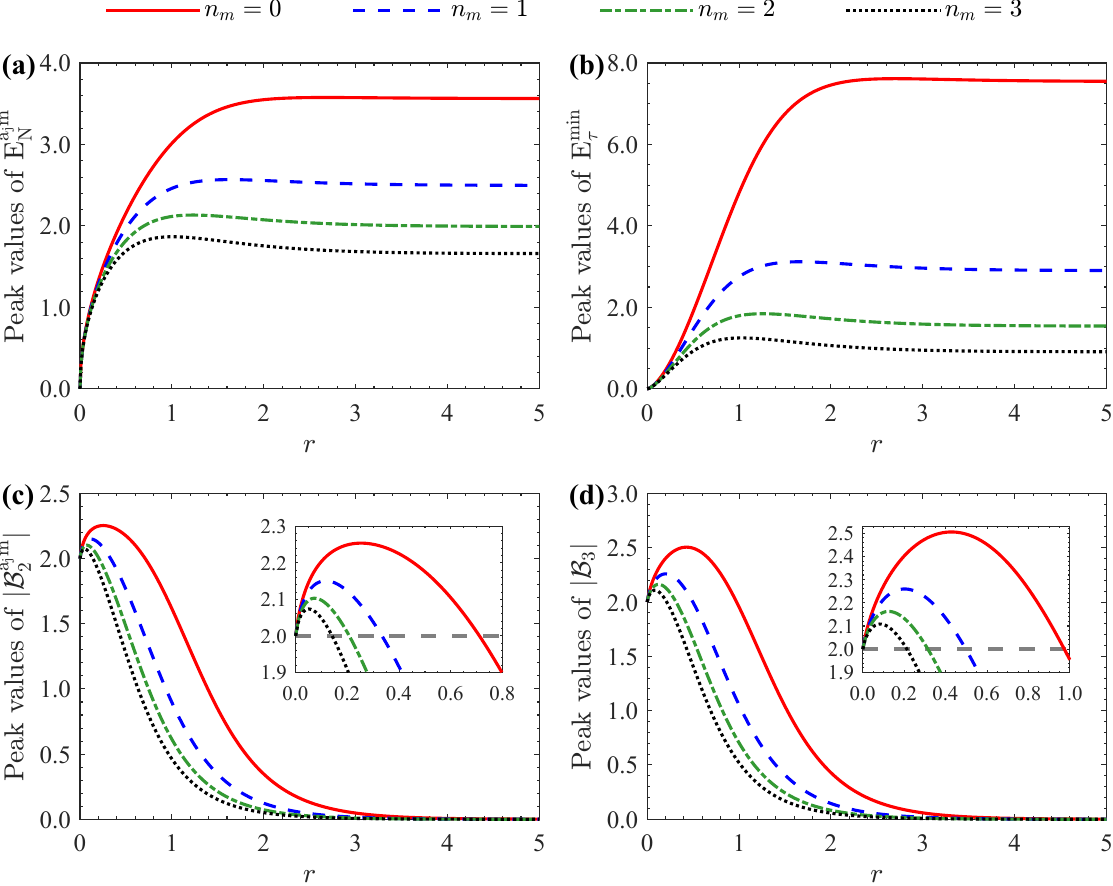}
\caption{The peak values of (a) the bipartite entanglement $E_N^{a_1 m}$ and $E_N^{a_2 m}$, (b) the genuine tripartite entanglement $E_{\tau}^{\min}$, (c) the bipartite Bell functions $|\mathcal{B}_2^{a_1 m}|$ and $|\mathcal{B}_2^{a_2 m}|$, and (d) the tripartite Bell function $|\mathcal{B}_3|$ versus the squeezing parameter $r$ with the relative phase $\psi$ optimized. 
The mechanical dissipation rate is set to $\gamma = 0.005g$, and the thermal noise occupation numbers are varied as $n_m = 0, 1, 2, 3$. The other parameters are the same as in Fig.~\ref{Fig_FunctionPsi}.}
\label{Fig_FunctionR_Loss}
\end{figure*}

It is evident from the preceding subsections that, in the ideal limit of negligible mechanical dissipation and noise, the peak values of both entanglement and Bell functions presented in Fig.~\ref{Fig_FunctionPsi} increase monotonically with the squeezing parameter $r$. However, the maximal bipartite Bell function remains substantially below the theoretical quantum-mechanical bound $2\sqrt{2}$, and the tripartite counterpart similarly falls far below the allowed value of $4$. A key question thus arises whether these ultimate Bell violations can be achieved and, if so, what value of the squeezing parameter $r$ is required. In this section, we address this question by investigating the lossless and noiseless regime in detail.

Figure~\ref{Fig_FunctionR_NoLoss} displays the maximal entanglement and Bell functions as functions of the squeezing parameter $r$, with the relative phase $\psi$ optimized for each quantity.
It is observed from Fig.~\ref{Fig_FunctionR_NoLoss}(a) that the maximal bipartite and tripartite entanglement exhibits monotonic growth without saturation across the entire range of $r$ considered.
In contrast to the monotonic growth of the entanglement measures, the bipartite Bell functions $|\mathcal{B}_2^{a_1 m}|$ and $|\mathcal{B}_2^{a_2 m}|$ along with the tripartite Bell function $|\mathcal{B}_3|$ exhibit an initial increase with $r$ followed by saturation, as depicted in Fig.~\ref{Fig_FunctionR_NoLoss}(b). 
For sufficiently large $r$, the bipartite Bell functions saturate at approximately $2.32$, which remains substantially below the quantum-mechanical maximum $2\sqrt{2}$. Notably, this saturation value  $2.32$ coincides with the maximal Bell-inequality violation attainable by an ideal two-mode squeezed, i.e., Einstein-Podolsky-Rosen (EPR), state under the displaced‑parity measurement~\cite{Ferraro2005_JOptB_BiTriBell,Jeong2003PRA_MultipartiteBell}. 
Similarly, the tripartite Bell function $|\mathcal{B}_3|$ converges to $3$ as $r$ increases, significantly lower than the theoretical maximum of $4$, which is the corresponding maximal violation attainable for an ideal tripartite EPR state under the displaced‑parity measurement~\cite{Ferraro2005_JOptB_BiTriBell}.
The underlying reason why the bipartite and tripartite Bell nonlocality cannot attain their maximal violation stems from the intrinsic limitation of the displaced‑parity measurement. Specifically, the displaced‑parity operator defined following Eq.~(\ref{WignerFunction}) does not guarantee a complete parity inversion, which restricts the two‑mode squeezed state from reaching the ultimate Bell‑inequality violation~\cite{Jeong2003PRA_MultipartiteBell}.
If alternative schemes such as pseudospin‑based Bell tests were employed, the bipartite violation could, in principle, reach the quantum limit~\cite{Ferraro2005_JOptB_BiTriBell,ChenZeng_Bing2002PRL_BellTest_Pseudospin}. However, pseudospin measurements are experimentally more challenging to implement. 
We thus adopt displaced-parity measurements as a compromise between Bell-violation strength and experimental accessibility.

\subsection{Effects of the mechanical dissipation $\gamma$ and noise $n_m$ on entanglement and Bell nonlocality}
\label{Results_Influence_DissipationLoss}

\begin{figure*}
\centering
\includegraphics[width=1.5\columnwidth]{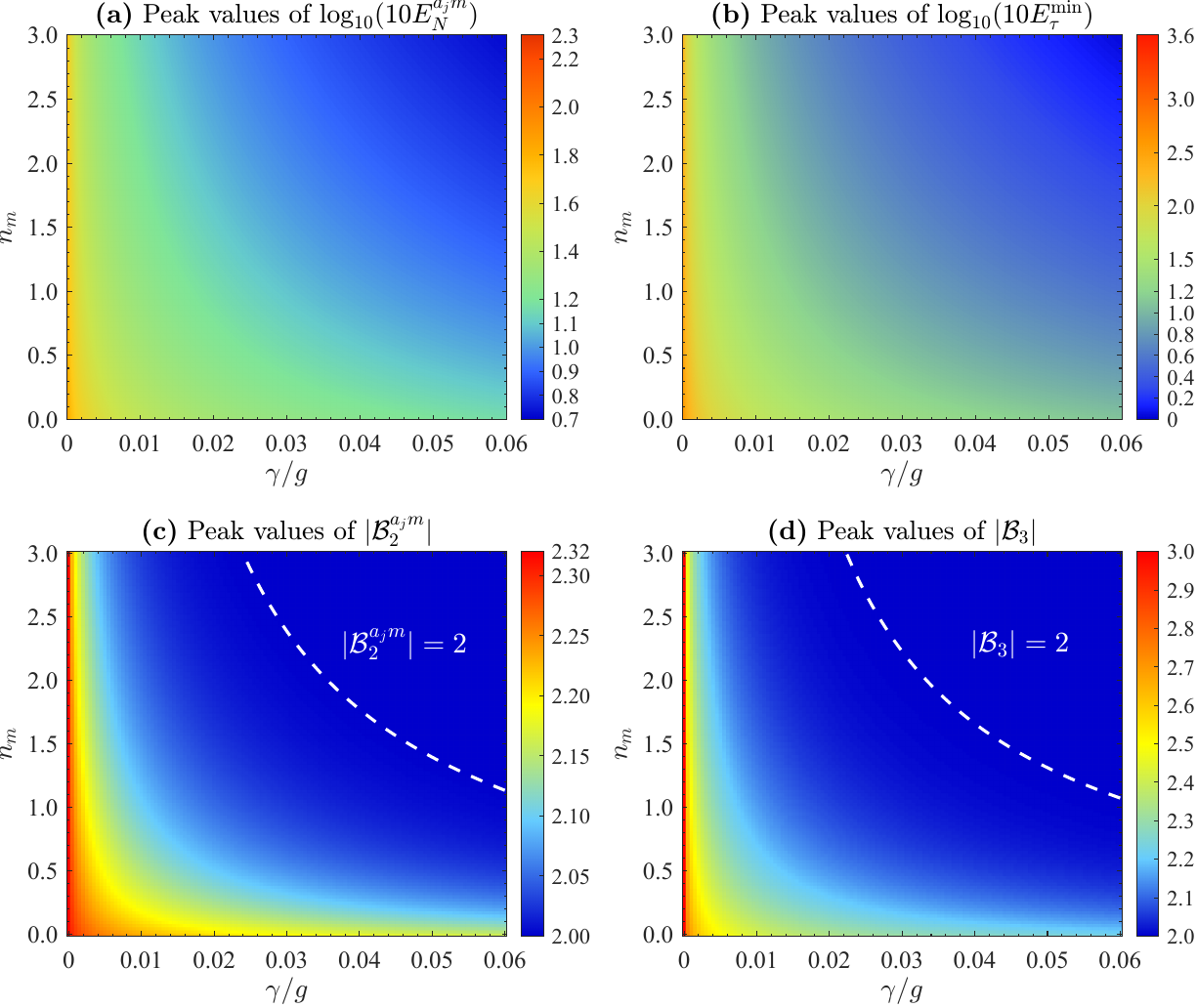}
\caption{The peak values of (a) the bipartite entanglement $E_N^{a_1 m}$ and $E_N^{a_2 m}$, (b) the genuine tripartite entanglement $E_{\tau}^{\min}$, (c) the bipartite Bell functions $|\mathcal{B}_2^{a_1 m}|$ and $|\mathcal{B}_2^{a_2 m}|$ and (d) the tripartite Bell function $|\mathcal{B}_3|$ as functions of the mechanical dissipation $\gamma$ and thermal noise $n_m$ by optimizing the squeezing factor $r$ and relative phase $\psi$. The other parameters are the same as in Fig.~\ref{Fig_FunctionPsi}.}
\label{Fig_Function_Loss_ThermalNoise}
\end{figure*}

In the previous subsections, we have investigated bipartite and tripartite entanglement together with Bell nonlocality under ideal conditions where the mechanical dissipation and noise are absent. In any realistic implementation, however, the mechanical oscillator inevitably interacts with its environment, introducing dissipation and thermal noise that can significantly degrade these quantum correlations. Therefore, in the following subsection, we systematically examine how the mechanical dissipation and thermal noise affect both entanglement and Bell nonlocality in the proposed closed-loop optomechanical system.

Figure~\ref{Fig_FunctionR_Loss} illustrates the dependence of the bipartite and tripartite entanglement and Bell functions on the squeezing parameter $r$ for selected values of mechanical damping $\gamma$ and thermal noise $n_m$. It is evident that introducing finite $\gamma$ and $n_m$ breaks the monotonic growth observed in the ideal lossless case. Instead, all entanglement and Bell measures initially increase with $r$, reach a maximum, and then decline as $r$ is further increased. For the Bell functions, this behavior implies that the violation first strengthens, peaks at an optimal $r$, and eventually falls below the classical bound, thereby losing nonlocality.
Compared to the bipartite and tripartite entanglement, Bell nonlocality is more fragile against the environmental noise. For the chosen parameters, bipartite and tripartite entanglement persist over the entire range of $r$ considered, whereas the Bell violations occur only within a limited interval of $r$, as seen in Figs.~\ref{Fig_FunctionR_Loss}(b) and (d). Larger thermal noise $n_m$ narrows this violation window and simultaneously reduces the peak violation strength.
Quantitatively, for thermal occupation numbers $n_m = 0,1,2,3$, the largest squeezing parameters that still permit a bipartite Bell violation are $r \approx 0.71,~0.33, ~0.20,~0.14$, and the corresponding maximal violations $|\mathcal{B}_2|_{\max} \approx  2.25,~2.15,~2.10,~2.07$, respectively. For tripartite nonlocality, the corresponding thresholds are $r \approx 0.97,~0.49,~0.31,~0.21$, yielding maxima $|\mathcal{B}_3|_{\max} \approx 2.51,~2.26,~2.16,~2.11$. In the presence of dissipation and noise, the peak violations remain substantially below the achievable limits $2.32$ (bipartite) and $3$ (tripartite) in lossless systems.

The two‑dimensional plots in Fig.~\ref{Fig_Function_Loss_ThermalNoise} further elucidate the dependence of entanglement and Bell functions on $\gamma$ and $n_m$. We note that both entanglement and Bell functions are degraded by mechanical damping and thermal noise, yet the Bell functions exhibit a higher degree of sensitivity. Violation of the Bell inequalities persists over a broad range of $\gamma$ and $n_m$, as indicated by the regions in the bottom‑left corner within the dashed contours. This robustness demonstrates that the proposed displaced‑parity‑based Bell test remains experimentally feasible under realistic dissipation and noise conditions.

Finally, we discuss the feasibility of experimentally verifying Bell nonlocality in the proposed system. For an optical mode, the displaced parity measurement can be implemented indirectly through homodyne detection or photon-number-resolving detection. In contrast, performing such a measurement directly on a mechanical mode  poses substantial experimental difficulties. Typically, the mechanical state can first be mapped onto an optical field, for instance through the optomechanical interaction, after which the displaced parity measurement is carried out on the resulting optical field~\cite{Vitali2007PRL_OptoEnt,Palomaki710}.

\section{Conclusion}\label{Conclusion}
In summary, we have studied the generation and phase-controlled manipulation of bipartite and tripartite entanglement and Bell nonlocality in a closed-loop three-mode optomechanical system. Two optical modes are coupled to a mechanical resonator via two-mode squeezing interactions induced by blue-detuned laser pulses, and are also directly coupled to each other through field transmission, forming a phase-sensitive cyclic configuration. The relative phase governs the population distribution of the optical modes, thereby enabling coherent redistribution of quantum correlations. 
By tuning this phase, the entanglement and Bell violations can be deterministically switched. At specific phase values, one optical-mechanical pair exhibits maximal bipartite entanglement and Bell violation, whereas at other phase values, the opposite pair becomes dominant. At symmetric points where the two bipartite entanglement measures are equal, the genuine tripartite entanglement and the tripartite Bell violation reach their respective maxima.

Under ideal lossless conditions, increasing the squeezing parameter enhances both bipartite and tripartite entanglement monotonically, while the maximal bipartite and tripartite Bell violations under displaced-parity measurements saturate at approximately $2.32$ and $3$, respectively. These values are the limiting violations attainable with the displaced-parity Bell test and remain below the corresponding quantum limits.
Introducing dissipation and thermal noise breaks this monotonicity. 
The entanglement measures and Bell functions first increase and then decrease, with Bell nonlocality being more sensitive to environmental dissipation and noise. Nevertheless, the Bell-inequality violations persist over broad parameter ranges, confirming the experimental feasibility under realistic conditions.

Last but not least, we demonstrate that violation of the tripartite Mermin-Klyshko-type Bell inequality employed here can persist even when the genuine tripartite entanglement vanishes.
Our results not only clarify the distinct roles of genuine tripartite entanglement and tripartite Bell nonlocality, but also provide a versatile platform for generating switchable bipartite and tripartite quantum correlations. Such phase‑controlled resources are essential for reconfigurable quantum networks and multiparty quantum information tasks, including device‑independent protocols and noise‑resilient tests of quantum foundations.

\begin{acknowledgements}
S. S. Zheng acknowledges the financial support from National Natural Science Foundation of China (Grant No.~12204441). F. X. Sun acknowledges National Natural Science Foundation of China  (Grant No.~12474256), the Quantum Science and Technology-National Science and Technology Major Project (Grant No. 2025ZD0301000), and the Fundamental Research Funds for the Central Universities (Grant No. 539926010). Q. Y. He acknowledges National Natural Science Foundation of China (Grant Nos. 12125402, 12534016), the Beijing Natural Science Foundation (Grant No. Z240007), and the Quantum Science and Technology-National Science and Technology Major Project (Grant Nos. 2024ZD0302401, 2021ZD0301500).
\end{acknowledgements}

\appendix\label{APPENDIX}

\section{Analytical solutions for bipartite/tripartite entanglement and Bell nonlocality at specific phases in the proposed closed-loop optomechanical system}\label{AppendixA}
Here, we analytically derive the covariance matrix for the parameter set employed in Fig.~\ref{Fig_FunctionPsi} and obtain exact expressions for the entanglement measures and Bell functions at three characteristic phase values, namely $\psi = (n + 3)\pi/4$, $\psi = (n+1) \pi/4$, and $\psi = n\pi/2$ (with $n = 0,1,2,\dots$). These phases correspond respectively to the regime where only the $a_1\!-\!m$ pair is entangled, the regime where only the $a_2\!-\!m$ pair is entangled, and the symmetric point where genuine tripartite entanglement is maximized. The resulting solutions illustrate how the relative phase $\psi$ coherently redistributes quantum correlations among the three modes.
\subsection{The phase $\psi = (n + 3/4)\pi$ ($n=0,1,2,\dots$)} \label{AppendixA_Psi0d75}
Firstly, when the phase $\psi = (n + 3/4)\pi$ ($n=0,1,2,\dots$), the covariance matrix of the full closed‑loop system reduces to the simple form
\begin{equation}
V=\left (
\begin{array}{cccccc}
x_0 & 0 & 0 & 0 & 0 & -x_1 \\
0 & x_0  & 0 & 0 & -x_1 & 0 \\
0 & 0 & 1/2 & 0  & 0 & 0 \\
0 & 0 &  0 & 1/2 & 0 & 0  \\
0 & -x_1 & 0 & 0 &  x_0 & 0 \\
-x_1 & 0 & 0 & 0 & 0 & x_0
\end{array} \right ),
\label{Appendix_TriPartiteState_CovarianceMatrix_Psi0d75}
\end{equation}
with $x_0=e^{2r}-1/2,~x_1=e^r\sqrt{e^{2r}-1}$. 
For any $r > 0$, we have $x_0 > 1/2$ and $0 < x_0 - x_1 < 1/2$.
In this appendix, for notational convenience, the three modes $a_1$, $a_2$, and $m$ are labeled as $1$, $2$, and $3$, respectively
According to the definitions of bipartite and tripartite entanglement introduced in Eqs.~(\ref{QuanMagnonics_LogNegativity}) and (\ref{Eq_QuanMagnonics_MinResContangle}) of the main manuscript, the relevant minimum symplectic eigenvalues are found to be 
\begin{eqnarray}
(\eta^-)^{12}&=&\min~\textrm{eig}|i \Omega P V_{\rm{reduced}}^{12} P|=1/2, \nonumber \\
(\eta^-)^{13}&=&\min~\textrm{eig}|i \Omega P V_{\rm{reduced}}^{13} P|=x_0-x_1,  \nonumber \\
(\eta^-)^{23}&=&\min~\textrm{eig}|i \Omega P V_{\rm{reduced}}^{23} P|=1/2,  \nonumber \\
(\eta^-)^{1|23}&=&\min~\textrm{eig}|i \Omega_3 P^{1|23} V P^{1|23}|=x_0-x_1,  \nonumber \\
(\eta^-)^{2|13}&=&\min~\textrm{eig}|i \Omega_3 P^{2|13} V P^{2|13}| =1/2, \nonumber \\
(\eta^-)^{3|12}&=&\min~\textrm{eig}|i \Omega_3 P^{3|12} V P^{3|12}|=x_0-x_1, \nonumber \\
\label{Appendix_TriPartiteState_MinimumSymplecticEigenvalue_Psi0d75}
\end{eqnarray}
where $(\eta^-)^{jk}$ represents the minimum symplectic eigenvalues of the one-to-one bipartite subsystems $j\!-\!k$, while $(\eta^-)^{j|kl}$ corresponds to the one-to-two bipartite subsystems $j\!-\!kl$ ($j,~k,~l=1,~2,~3$). And $V_{\rm{reduced}}^{jk}$ denotes the reduced covariance matrix of the $j-k$ subsystem. The matrices $\Omega=\bigoplus_{i=1}^2 i\hat{\sigma}_y,~P=\mathrm{diag}(1,-1,1,1),~\Omega_3=\bigoplus_{i=1}^3 i\hat{\sigma}_y,~P^{j \mid k l} =\mathrm{diag}(1,-1,1,1,1,1),~P^{k \mid j l} =\mathrm{diag}(1,1,1,-1,1,1),~P^{l \mid j k} =\mathrm{diag}(1,1,1,1,1,-1)~(j,~k,~l=1,~2,~3)$, which can be referred to Eqs.~(\ref{QuanMagnonics_LogNegativity}) and (\ref{Eq_QuanMagnonics_MinResContangle}) in the main manuscript for details.  

Therefore, the bipartite entanglement of the one-to-one bipartite subsystems reads
\begin{eqnarray}
E_N^{12}&=&\max\left\{0,~-\ln\left(2\left(\eta^-\right)^{12}\right)\right\}=0, \nonumber \\
E_N^{13}&=&\max\left\{0,~-\ln\left(2\left(\eta^-\right)^{13}\right)\right\}=-\ln\left[2\left(x_0-x_1\right)\right]\nonumber \\
&=&2\ln\left(e^{r}+\sqrt{e^{2r}-1}\right),  \nonumber \\
E_N^{23}&=&\max\left\{0,~-\ln\left(2\left(\eta^-\right)^{23}\right)\right\}=0.
\label{Appendix_TriPartiteState_OneToOneEnt_Psi0d75}
\end{eqnarray}
Note that $E_N^{13}$ grows monotonically with increasing squeezing parameter $r$.
And the bipartite entanglement of the one-to-two bipartite subsystems is quantified by 
\begin{eqnarray}
E_N^{1|23}&=&\max\left\{0,~-\ln\left(2\left(\eta^-\right)^{1|23}\right)\right\}=-\ln\left[2\left(x_0-x_1\right)\right] \nonumber \\
&=&2\ln\left(e^{r}+\sqrt{e^{2r}-1}\right),  \nonumber \\
E_N^{2|13}&=&\max\left\{0,~-\ln\left(2\left(\eta^-\right)^{2|13}\right)\right\}=0, \nonumber \\
E_N^{3|12}&=&\max\left\{0,~-\ln\left(2\left(\eta^-\right)^{3|12}\right)\right\}=-\ln\left[2\left(x_0-x_1\right)\right]\nonumber \\
&=&2\ln\left(e^{r}+\sqrt{e^{2r}-1}\right). 
\label{Appendix_TriPartiteState_OneToTwoEnt_Psi0d75}
\end{eqnarray}

The residual contangle defined in Eq.~(\ref{Eq_QuanMagnonics_ResContangle}) of the main manuscript aiming for computing the genuine tripartite entanglement is given by  
\begin{eqnarray}
E_{\tau}^{1|23}&=&C_{\tau}^{1|23}-C_{\tau}^{1|2}-C_{\tau}^{1|3} \nonumber \\
&=&(E_{N}^{1|23})^2-(E_{N}^{12})^2-(E_{N}^{13})^2=0, \nonumber \\
E_{\tau}^{2|13}&=&C_{\tau}^{2|13}-C_{\tau}^{2|1}-C_{\tau}^{2|3} \nonumber \\
&=&(E_{N}^{2|13})^2-(E_{N}^{12})^2-(E_{N}^{23})^2=0, \nonumber \\
E_{\tau}^{3|12}&=&C_{\tau}^{3|12}-C_{\tau}^{3|1}-C_{\tau}^{3|2} \nonumber \\ 
&=&(E_{N}^{3|12})^2-(E_{N}^{13})^2-(E_{N}^{23})^2=0.
\label{Appendix_TriPartiteState_ResidualContangle_Psi0d75}
\end{eqnarray}
Therefore, the genuine tripartite entanglement reads
\begin{eqnarray}
E_{\tau}^{\min}=\min\left\{E_{\tau}^{1|23},~E_{\tau}^{2|13},~E_{\tau}^{3|12}\right\}=0.
\label{Appendix_TriPartiteState_MinimalResidualContangle_Psi0d75}
\end{eqnarray}

For the bipartite and tripartite Bell functions, the Wigner functions can be derived through symbolic calculations. The resulting expression is too lengthy to be presented here. Instead, the relationship between the Wigner functions of the bipartite and tripartite systems takes the simple form
\begin{eqnarray}
W_{123}(u_1,~u_2,~u_3)=\frac{2}{\pi}\exp\left[-2|u_2|^2\right]W_{13}\left(u_1,~u_3\right). 
\label{Appendix_TriPartiteState_WignerBiTri_Psi0d75}
\end{eqnarray} 
Consequently, the tripartite Bell function can be formulated as
\begin{eqnarray}
\mathcal{B}_{123}&=&\frac{\pi^3}{8}\Big[W_{123}(u_1,~u_2,~u_3’)+W_{123}(u_1,~u_2’,~u_3)  \nonumber \\
&&+W_{123}(u_1’,~u_2,~u_3)-W_{123}(u_1’,~u_2’,~u_3’)\Big] \nonumber \\
&=&\frac{\pi^2}{4}\Big\{\exp\left[-2|u_2|^2\right]W_{13}\left(u_1,~u_3’\right) \nonumber \\
&&+\exp\left[-2|u_2’|^2\right]W_{13}\left(u_1,~u_3\right) \nonumber \\
&&+\exp\left[-2|u_2|^2\right]W_{13}\left(u_1’,~u_3\right) \nonumber \\
&&-\exp\left[-2|u_2’|^2\right]W_{13}\left(u_1’,~u_3’\right)\Big\}.
\label{Appendix_TriPartiteState_BellFuncBiTri_Psi0d75}
\end{eqnarray} 
Taking $u_2=u_2’=0$, we have
\begin{eqnarray}
\mathcal{B}_{123}&=&\frac{\pi^2}{4}\Big[W_{13}\left(u_1,~u_3\right)+W_{13}\left(u_1,~u_3’\right)+W_{13}\left(u_1’,~u_3\right) \nonumber\\
&&-W_{13}\left(u_1’,~u_3’\right)\Big], \nonumber \\
&=&\mathcal{B}_{13}.
\label{Appendix_TriPartiteState_BellFuncBiTriOpt_Psi0d75}
\end{eqnarray} 
The above expression indicates that for the phase $\psi = (n+ 3/4)\pi$ ($n=0,1,2,\dots$), the tripartite Bell function of our closed-loop three-mode system reduces exactly to the bipartite Bell function of the $a_1\!-\!m$ two-mode subsystem. 

\subsection{The phase $\psi = (n+ 1/4)\pi $ ($n=0,1,2,\dots$)} \label{AppendixA_Psi0d25}
When the phase takes the value $\psi = (n+ 1/4)\pi $ ($n=0,1,2,\dots$), the covariance matrix simplifies to 
\begin{equation}
V=\left (
\begin{array}{cccccc}
1/2 & 0 & 0 & 0 & 0 & 0 \\
0 & 1/2  & 0 & 0 & 0 & 0 \\
0 & 0 & x_0 & 0  & 0 & -x_1 \\
0 & 0 &  0 & x_0 & -x_1 & 0  \\
0 & 0& 0 & -x_1 &  x_0 & 0 \\
0 & 0 & -x_1 & 0 & 0 & x_0
\end{array} \right ).
\label{Appendix_TriPartiteState_CovarianceMatrix_Psi0d25}
\end{equation}

Comparing the covariance matrix Eq.~(\ref{Appendix_TriPartiteState_CovarianceMatrix_Psi0d25}) when $\psi = (n+ 1/4)\pi $ ($n=0,1,2,\dots$) with Eq.~(\ref{Appendix_TriPartiteState_CovarianceMatrix_Psi0d75}) when $\psi = (n+ 3/4)\pi $ ($n=0,1,2,\dots$), we note that the role of $a_1$ and $a_2$ are interchanged. Following the derivation in Appendix~\ref{AppendixA_Psi0d75}, the bipartite entanglement are obtained 
\begin{eqnarray}
E_N^{12}&=&\max\left\{0,~-\ln(2(\eta^-)^{12})\right\}=0, \nonumber \\
E_N^{13}&=&\max\left\{0,~-\ln(2(\eta^-)^{13})\right\}=0, \nonumber \\
E_N^{23}&=&\max\left\{0,~-\ln(2(\eta^-)^{23})\right\}=-\ln\left[2(x_0-x_1)\right]\nonumber \\
&=&2\ln\left(e^{r}+\sqrt{e^{2r}-1}\right).  
\label{Appendix_TriPartiteState_OneToOneEnt_Psi0d25}
\end{eqnarray}
And the bipartite entanglement of the one-to-two bipartite subsystems can be quantified by 
\begin{eqnarray}
E_N^{1|23}&=&\max\left\{0,~-\ln(2(\eta^-)^{1|23})\right\}=0, \nonumber \\
E_N^{2|13}&=&\max\left\{0,~-\ln(2(\eta^-)^{2|13})\right\}=-\ln\left[2(x_0-x_1)\right] \nonumber \\
&=&2\ln\left(e^{r}+\sqrt{e^{2r}-1}\right).  \nonumber \\
E_N^{3|12}&=&\max\left\{0,~-\ln(2(\eta^-)^{3|12})\right\}=-\ln\left[2(x_0-x_1)\right]\nonumber \\
&=&2\ln\left(e^{r}+\sqrt{e^{2r}-1}\right). 
\label{Appendix_TriPartiteState_OneToTwoEnt_Psi0d25}
\end{eqnarray}
Consequently, the genuine tripartite entanglement reads
\begin{eqnarray}
E_{\tau}^{\min}=\min\left\{E_{\tau}^{1|23},~E_{\tau}^{2|13},~E_{\tau}^{3|12}\right\}=0,
\label{Appendix_TriPartiteState_MinimalResidualContangle_Psi0d25}
\end{eqnarray}
indicating the absence of genuine tripartite entanglement at this phase.

Similarly, the tripartite Bell function is reduced to the bipartite Bell function
\begin{eqnarray}
\mathcal{B}_{123}&=&\frac{\pi^2}{4}\Big[W_{23}\left(u_2,~u_3\right)+W_{23}\left(u_2,~u_3’\right)+W_{23}\left(u_2’,~u_3\right) \nonumber\\
&&-W_{13}\left(u_2’,~u_3’\right)\Big], \nonumber \\
&=&\mathcal{B}_{23}.
\label{Appendix_TriPartiteState_BellFuncBiTriOpt_Psi0d25}
\end{eqnarray} 

\subsection{The phase $\psi = n\pi/2$ ($n=0,1,2,\dots$)} \label{AppendixA_Psi0}
When the phase $\psi = n\pi/2$ ($n=0,1,2,\dots$), the covariance matrix of the whole closed-loop system takes the form of 
\begin{equation}
V=\left (
\begin{array}{cccccc}
\frac{2x_0+1}{4} & 0  & \frac{2x_0-1}{4} & 0 & 0 & -\frac{x_1}{\sqrt{2}}\\
0 & \frac{2x_0+1}{4}  & 0 & \frac{2x_0-1}{4} & -\frac{x_1}{\sqrt{2}} & 0\\
\frac{2x_0-1}{4} & 0 &  \frac{2x_0+1}{4} & 0 & 0 & -\frac{x_1}{\sqrt{2}}\\
0 & \frac{2x_0-1}{4} & 0 & \frac{2x_0+1}{4} & -\frac{x_1}{\sqrt{2}} & 0\\
0 & -\frac{x_1}{\sqrt{2}} & 0 & -\frac{x_1}{\sqrt{2}} & x_0 & 0\\
-\frac{x_1}{\sqrt{2}} & 0 & -\frac{x_1}{\sqrt{2}} & 0 &  0 & x_0
\end{array} \right ).
\label{Appendix_TriPartiteState_CovarianceMatrix_Psi0}
\end{equation}
The relevant minimum symplectic eigenvalues are respectively
\begin{eqnarray}
(\eta^-)^{12}&=&\min~\textrm{eig}\left|i \Omega P V_{\rm{reduced}}^{12} P\right| \nonumber \\
&=&\frac{\sqrt{2e^{2r}-1}}{2}, \nonumber \\
(\eta^-)^{13}&=&\min~\textrm{eig}\left|i \Omega P V_{\rm{reduced}}^{13} P\right|  \nonumber \\
&=&\frac{1}{3-e^{-2r}+\sqrt{\left(1-e^{-2r}\right)\left(9-e^{-2r}\right)}}, \nonumber \\
(\eta^-)^{23}&=&\min~\textrm{eig}\left|i \Omega P V_{\rm{reduced}}^{23} P\right|  \nonumber \\
&=&\frac{1}{3-e^{-2r}+\sqrt{\left(1-e^{-2r}\right)\left(9-e^{-2r}\right)}}, \nonumber \\
(\eta^-)^{1|23}&=&\min~\textrm{eig}\left|i \Omega_3 P^{1|23} V P^{1|23}\right|  \nonumber \\
&=&\frac{e^{2r}-\sqrt{e^{4r}-1}}{2}, \nonumber \\
(\eta^-)^{2|13}&=&\min~\textrm{eig}\left|i \Omega_3 P^{2|13} V P^{2|13}\right|  \nonumber \\
&=&\frac{e^{2r}-\sqrt{e^{4r}-1}}{2}, \nonumber \\
(\eta^-)^{3|12}&=&\min~\textrm{eig}\left|i \Omega_3 P^{3|12} V P^{3|12}\right|  \nonumber \\
&=&\frac{\left(e^{r}-\sqrt{e^{2r}-1}\right)^2}{2}.
\label{Appendix_TriPartiteState_MinimumSymplecticEigenvalue_Psi0}
\end{eqnarray}
Therefore, the bipartite entanglement of the one-to-one bipartite subsystems reads
\begin{eqnarray}
E_N^{12}&=&\max\left\{0,~-\ln\left(2\left(\eta^-\right)^{12}\right)\right\}=0, \nonumber \\
E_N^{13}&=&\max\left\{0,~-\ln\left(2\left(\eta^-\right)^{13}\right)\right\} \nonumber \\
&=&\ln\left(\frac{3-e^{-2r}+\sqrt{\left(1-e^{-2r}\right)\left(9-e^{-2r}\right)}}{2}\right), \nonumber \\
E_N^{23}&=&\max\left\{0,~-\ln\left(2\left(\eta^-\right)^{23}\right)\right\} \nonumber\\
&=&\ln\left(\frac{3-e^{-2r}+\sqrt{\left(1-e^{-2r}\right)\left(9-e^{-2r}\right)}}{2}\right). \nonumber \\
\label{Appendix_TriPartiteState_OneToOneEnt_Psi0}
\end{eqnarray}
It is easily found that $E_N^{13}=E_N^{23}\rightarrow -\ln\left(2/6\right)=1.0986$ for $r\rightarrow \infty$.
And the bipartite entanglement of the one-to-two bipartite subsystems are 
\begin{eqnarray}
E_N^{1|23}&=&\max\left\{0,~-\ln\left(2\left(\eta^-\right)^{1|23}\right)\right\} \nonumber \\
&=&\ln\left(e^{2r}+\sqrt{e^{4r}-1}\right), \nonumber \\
E_N^{2|13}&=&\max\left\{0,~-\ln\left(2\left(\eta^-\right)^{2|13}\right)\right\} \nonumber \\
&=&\ln\left(e^{2r}+\sqrt{e^{4r}-1}\right), \nonumber \\
E_N^{3|12}&=&\max\left\{0,~-\ln\left(2\left(\eta^-\right)^{3|12}\right)\right\} \nonumber \\
&=&2\ln\left(e^{r}+\sqrt{e^{2r}-1}\right).
\label{Appendix_TriPartiteState_OneToTwoEnt_Psi0}
\end{eqnarray}
Therefore, after evaluating the residual contangle defined in Eq.~(\ref{Eq_QuanMagnonics_ResContangle}), the genuine tripartite entanglement takes the form
\begin{eqnarray}
E_{\tau}^{123}&=&\min\left\{E_{\tau}^{1|23},~E_{\tau}^{2|13},~E_{\tau}^{3|12}\right\}=E_{\tau}^{3|12}\nonumber\\
&=&4\left[ \ln\left(e^{r}+\sqrt{e^{2r}-1}\right)\right]^2 \nonumber\\
&&-2\left[ \ln\left(\frac{3-e^{-2r}+\sqrt{(1-e^{-2r})(9-e^{-2r})}}{2}\right)\right]^2. \nonumber\\
\label{Appendix_TriPartiteState_MinimalResidualContangle_Psi0}
\end{eqnarray}
Both terms grow with increasing $r$, however, the second term grows more slowly and satuarates when $r\rightarrow \infty$. Consequently, the genuine tripartite entanglement $E_{\tau}^{123}$ increases monotonically with $r$.

\section{Bipartite/tripartite entanglement and Bell nonlocality in an artificially constructed three-mode system}
\label{Appendix_ArtificalThreeModes}
To better understand the relationship between genuine tripartite entanglement and tripartite Bell nonlocality, we artificially construct a simplified three-mode system.
Through the detailed analysis of this artificial three-mode system, one can easily conclude that the genuine tripartite entanglement is not a prerequisite for the violation of tripartite Bell nonlocality. Even when the genuine tripartite entanglement is completely absent, the tripartite Bell inequality can still be violated significantly.

Here we first introduce the artificial three-mode system and provide the corresponding covariance matrix, which is essential for analyzing entanglement and Bell nonlocality. In this model, modes $A$ and $B$ form a standard two-mode squeezed state, described by $|S_{AB}\rangle=\exp{[r(A^\dagger B^\dagger - AB)]}|0\rangle_A \otimes |0\rangle_B$ with $r$ being the squeezing parameter, which is a positive real number. Besides, the third mode $C$ is completely separable from two modes $A$ and $B$, remaining in its vacuum state. Consequently, the state of the above three-mode system takes the form of 
\begin{equation}
|S_{\rm{tri}}\rangle=\exp{[r(A^\dagger B^\dagger - AB)]}|0\rangle_A \otimes |0\rangle_B \otimes |0\rangle_C.
\label{Artifical_TriPartiteState}
\end{equation}
Given the description of the covariance matrix of the two-mode squeezed state~\cite{Yuan_2020PRB_MagnonEnt,Gerardo_2014_CVSystem}, we obtain the covariance matrix of the artificial three‑mode system as
\begin{equation}
V_{\rm{tri}}=\left (
\begin{array}{cccccc}
 n & 0 & m & 0 & 0 & 0\\
0 & n & 0 & -m & 0 & 0\\
m & 0 & n & 0 & 0 & 0\\
0 & -m & 0 & n & 0 & 0\\
0 & 0 & 0 & 0 & 1/2 & 0\\
0 & 0 & 0 & 0 & 0 & 1/2
\end{array} \right ),
\label{Artifical_TriPartiteState_CovarianceMatrix}
\end{equation}
where $n=\cosh{2r}/2$ and $m=\sinh{2r}/2$. 
Here, the definition of the covariance matrix follows the same convention as in Eq.~(\ref{Eq_CovarianceMatrix}) of the main manuscript.

\subsection{Bipartite and tripartite entanglement}
\label{Appendix_ArtificalThreeModes_Ent}
With the covariance matrix of our artificial three-mode system given in  Eq.~(\ref{Artifical_TriPartiteState_CovarianceMatrix}), we now discuss the properties of both the bipartite and tripartite entanglement. The system possesses three one-to-one bipartite subsystems (denoted as 1-2, 1-3 and 2-3 subsystems) and three one-to-two bipartite subsystems (denoted as 1-23, 2-13 and 3-12 subsystems). Here, for notational simplicity, the three modes $A,~B,~C$ are labeled as $1,2,3$, respectively.  

Using the definitions of bipartite and tripartite entanglement introduced in Eqs.~\eqref{QuanMagnonics_LogNegativity} and \eqref{Eq_QuanMagnonics_MinResContangle} of the main manuscript, the relevant minimum symplectic eigenvalues are found to be
\begin{eqnarray}
(\eta^-)^{12}&=&\min~\textrm{eig}|i \Omega P V_{\rm{tri}}^{12} P|=e^{-2r}/2, \nonumber \\
(\eta^-)^{13}&=&\min~\textrm{eig}|i \Omega P V_{\rm{tri}}^{13} P|=1/2 , \nonumber \\
(\eta^-)^{23}&=&\min~\textrm{eig}|i \Omega P V_{\rm{tri}}^{23} P| =1/2, \nonumber \\
(\eta^-)^{1|23}&=&\min~\textrm{eig}|i \Omega_3 P^{1|23} V_{\rm{tri}} P^{1|23}|=e^{-2r}/2,  \nonumber \\
(\eta^-)^{2|13}&=&\min~\textrm{eig}|i \Omega_3 P^{2|13} V_{\rm{tri}} P^{2|13}|=e^{-2r}/2 , \nonumber \\
(\eta^-)^{3|12}&=&\min~\textrm{eig}|i \Omega_3 P^{3|12} V_{\rm{tri}} P^{3|12}|=1/2, 
\label{Artifical_TriPartiteState_MinimumSymplecticEigenvalue}
\end{eqnarray}
where $V_{\rm{tri}}^{jk}$ denotes the reduced covariance matrix of the $j$‑$k$ subsystem. The matrices $\Omega$, $P$, $\Omega_3$, and $P^{j|kl}$ are defined in the main text (see Eqs.~\eqref{QuanMagnonics_LogNegativity} and \eqref{Eq_QuanMagnonics_MinResContangle} for details).

\begin{figure*}
\centering
\includegraphics[width=2\columnwidth]{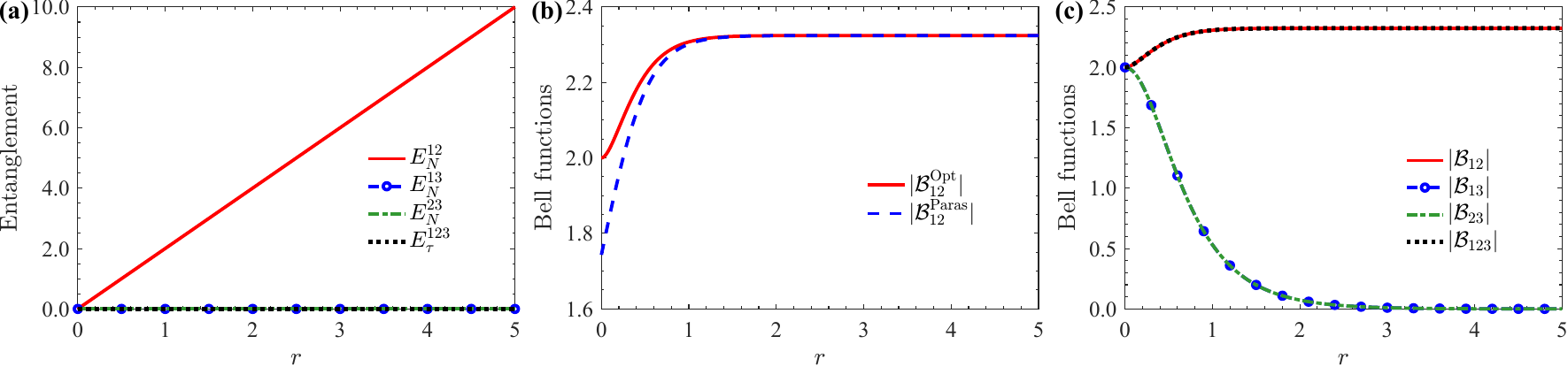}
\caption{(a) Bipartite and tripartite entanglement and (b-c) the corresponding Bell functions versus the squeezing parameter $r$. Figure~(b) demonstrates the Bell function of the 1-2 subsystem, in which the red solid curve indicates the maximized value of an absolute Bell function $|\mathcal{B}_{12}|_{\max}$ with increasing $r$ through numerical optimization, while the blue dashed-dotted curve displays $|\mathcal{B}_{12}|$ through theoretical calculation when choosing specific optimal parameters $u_1,u_2,u_1’,u_2’$ defined in Eq.~(\ref{Artifical_TriPartiteState_BipartiteBell12_OptParas}). }
\label{Fig_ArtificialThreeMode}
\end{figure*}

Therefore, the bipartite entanglement of the one-to-one bipartite subsystems reads
\begin{eqnarray}
E_N^{12}&=&\max\left\{0,~-\ln\left(2\left(\eta^-\right)^{12}\right)\right\}=2r, \nonumber \\
E_N^{13}&=&\max\left\{0,~-\ln\left(2\left(\eta^-\right)^{13}\right)\right\}=0, \nonumber \\
E_N^{23}&=&\max\left\{0,~-\ln\left(2\left(\eta^-\right)^{23}\right)\right\}=0,
\label{Artifical_TriPartiteState_OneToOneEnt}
\end{eqnarray}
and the bipartite entanglement of the one-to-two bipartite subsystems is 
\begin{eqnarray}
E_N^{1|23}&=&\max\left\{0,~-\ln\left(2\left(\eta^-\right)^{1|23}\right)\right\}=2r, \nonumber \\
E_N^{2|13}&=&\max\left\{0,~-\ln\left(2\left(\eta^-\right)^{2|13}\right)\right\}=2r, \nonumber \\
E_N^{3|12}&=&\max\left\{0,~-\ln\left(2\left(\eta^-\right)^{3|12}\right)\right\}=0.
\label{Artifical_TriPartiteState_OneToTwoEnt}
\end{eqnarray}
Using the residual contangle defined in Eq.~\eqref{Eq_QuanMagnonics_ResContangle}, the genuine tripartite entanglement reads
\begin{eqnarray}
E_{\tau}^{123}=\min\left\{E_{\tau}^{1|23},~E_{\tau}^{2|13},~E_{\tau}^{3|12}\right\}=0.
\label{Artifical_TriPartiteState_MinimalResidualContangle}
\end{eqnarray}

The above analytical results show that the artificial three-mode system possesses strong one-to-one entanglement $E_N^{12}=2r$ and one-to-two entanglement $E_N^{1|23}=E_N^{2|13}=2r$, while all other bipartite entanglement and the genuine  tripartite entanglement vanish. For verification, we plot both the bipartite and tripartite entanglement of this artificial three-mode system in Fig.~\ref{Fig_ArtificialThreeMode}(a).
The numerical results agree perfectly with the analytical expressions. Specifically, $E_N^{12}$ increases linearly with $r$, whereas $E_N^{13}$, $E_N^{23}$, and $E_{\tau}^{123}$ remain zero. In the following, we further investigate the bipartite and tripartite Bell nonlocality to determine whether tripartite Bell nonlocality can be achieved even in the absence of genuine tripartite entanglement.

\subsection{Bipartite and tripartite Bell nonlocality}
\label{Appendix_ArtificalThreeModes_Bell}
In the preceding subsection, we derived analytical formulas for both bipartite and tripartite entanglement in the artificial three‑mode system and presented numerical simulations of their behavior. We now turn to the analysis of bipartite and tripartite Bell nonlocality, with the aim of confirming that genuine tripartite entanglement is not a necessary condition for the violation of tripartite Bell nonlocality. As in the main text, Bell nonlocality is tested via displaced parity measurements.

Using the covariance matrix of the artificial system given in Eq.~(\ref{Artifical_TriPartiteState_CovarianceMatrix}), the Wigner functions for the various subsystems are obtained as
\begin{eqnarray}
W_{12}(u_1,u_2)&=&\frac{4}{\pi^2}\exp\Big[-2\cosh 2r (|u_1|^2+|u_2|^2) \nonumber \\
&&+ 2\sinh 2r (u_1u_2+u_1^* u_2^*)\Big],  \nonumber\\
W_{13}(u_1,u_2)&=&\frac{4}{\pi^2\cosh^2 r}\exp\left[-2 \left(\frac{|u_1|^2}{\cosh 2r}+|u_2|^2\right)\right], \nonumber \\
W_{23}(u_1,u_2)&=&\frac{4}{\pi^2\cosh^2 r}\exp\left[-2 \left(\frac{|u_1|^2}{\cosh 2r}+|u_2|^2\right)\right],\nonumber\\
\label{Artifical_TriPartiteState_BipartiteWigner}
\end{eqnarray}
where $W_{12}(u_1,u_2),~W_{13}(u_1,u_2)$ and $W_{23}(u_1,u_2)$ respectively represent the Wigner functions of the 1-2, 1-3 and 2-3 subsystems.
Note that the Wigner function $W_{12}$ of the 1-2 subsystem is in full agreement with previous works~\cite{LiJie_2017PRA_OptoBell, vanLoock_2001PRA_GHZBell,Barnett_1987JourModernOpt_Squeezing}. 

The corresponding Bell function for the 1-2 subsystem takes the form of 
\begin{eqnarray}
\mathcal{B}_{12}&=&\frac{\pi^2}{4}\Big[W_{12}\left(u_1,u_2\right)+W_{12}\left(u_1’,u_2\right)+W_{12}\left(u_1,u_2’\right)\nonumber\\
 &&-W_{12}\left(u_1’,u_2’\right)\Big] \nonumber\\
&=&\exp\Big[2\sinh 2r \left(u_1u_2+u_1^* u_2^*\right) \nonumber\\
&&~~~~~~~-2\cosh 2r \left(|u_1|^2+|u_2|^2\right)\Big] \nonumber\\
&&+\exp\Big[2\sinh 2r \left(u_1’u_2+u_1’^* u_2^*\right) \nonumber\\
&&~~~~~~~~~-2\cosh 2r \left(|u_1’|^2+|u_2|^2\right)\Big] \nonumber\\
&&+\exp\Big[2\sinh 2r \left(u_1u_2’+u_1^* u_2’^*\right) \nonumber\\
&&~~~~~~~~~-2\cosh 2r \left(|u_1|^2+|u_2’|^2\right)\Big] \nonumber\\
&&-\exp\Big[2\sinh 2r \left(u_1’u_2’+u_1’^* u_2’^*\right) \nonumber\\
&&~~~~~~~~~-2\cosh 2r \left(|u_1’|^2+|u_2’|^2\right)\Big],
\label{Artifical_TriPartiteState_BipartiteBell12}
\end{eqnarray}
According to Ref.~\cite{Jeong2003PRA_MultipartiteBell}, for a fixed squeezing parameter $r$, $|\mathcal{B}_{12}|$ takes the maximum when 
\begin{equation}
u_1=-u_1’=u_2’/2=\sqrt{\ln3/(16 \cosh 2r)},~u_2=0.
\label{Artifical_TriPartiteState_BipartiteBell12_OptParas}
\end{equation}
The variation of $|\mathcal{B}_{12}|$ versus the squeezing parameter $r$ adopting the above optimal parameters shown in Eq.~(\ref{Artifical_TriPartiteState_BipartiteBell12_OptParas}) is displayed in Fig.~\ref{Fig_ArtificialThreeMode}(b), manifested by the blue dash-dotted curve. For comparison, we plot the maximized value of $|\mathcal{B}_{12}|$ through numerical optimization of parameters $u_j$ and $u_j’~(j=1,2)$, displayed by the red solid curve. 
It can be clearly seen that the two curves agree very well for large $r$, nearly overlapping when $r>1.4$. The asymptotic violation of Bell's inequality is $|\mathcal{B}_{12}|_{\max}\approx 2.32$ for large squeezing, which is known to be the maximal violation attainable for a two‑mode squeezed state under displaced parity measurement~\cite{LiJie_2017PRA_OptoBell,Barnett_1987JourModernOpt_Squeezing}.

From Eq.~(\ref{Artifical_TriPartiteState_BipartiteWigner}), we have $W_{13}\left(u_1,~u_2\right)=W_{23}\left(u_1,~u_2\right)$, so the Bell functions for the 1‑3 and 2‑3 subsystems are identical
\begin{eqnarray}
\mathcal{B}_{13}=\mathcal{B}_{23}&=&\frac{\pi^2}{4}\Big[W_{23}\left(u_1,~u_2\right)+W_{23}\left(u_1’,~u_2\right) \nonumber \\
&&+W_{23}\left(u_1,~u_2’\right)-W_{23}\left(u_1’,~u_2’\right)\Big] \nonumber \\
&=&\frac{1}{\cosh ^2 r}\Big\{ \exp\left[-2 \left(|u_1|^2/\cosh 2r+|u_2|^2\right)\right] \nonumber \\
&&+\exp\left[-2 \left(|u_1’|^2/\cosh 2r+|u_2|^2\right)\right] \nonumber \\
&&+\exp\left[-2 \left(|u_1|^2/\cosh 2r+|u_2’|^2\right)\right] \nonumber \\
&&-\exp\left[-2 \left(|u_1’|^2/\cosh 2r+|u_2’|^2\right)\right]\Big \}, \nonumber \\
\label{Artifical_TriPartiteState_BipartiteBell13_23}
\end{eqnarray}
The maximum of $|\mathcal{B}_{13}|$ and $|\mathcal{B}_{23}|$ occur for $u_1=u_1'=u_2=0$ with arbitrary $u_2'$, yielding $|\mathcal{B}_{13}|_{\max}=|\mathcal{B}_{23}|_{\max}=2/\cosh^2 r <2$ for $r>0$. Hence, Bell inequality cannot be violated for the 1‑3 and 2‑3 subsystems through displaced parity measurement. This is confirmed by numerical optimization in Fig.~\ref{Fig_ArtificialThreeMode}(c), as displayed by the blue and green curves.

We now turn to study the tripartite Bell nonlocality. The Wigner function of the full three-mode system can be represented by 
\begin{eqnarray}
W_{123}(u_1,~u_2,~u_3)&=&\frac{8}{\pi^3}\exp\Big[-2\cosh 2r \left(|u_1|^2+|u_2|^2\right) \nonumber \\
&&+2\sinh 2r \left(u_1u_2+u_1^* u_2^*\right)-2|u_3|^2\Big], \nonumber \\
&=&\frac{2}{\pi}\exp\left[-2|u_3|^2\right]W_{12}\left(u_1,~u_2\right). \nonumber \\
\end{eqnarray} 
The tripartite Bell function can thus be formulated as
\begin{eqnarray}
\mathcal{B}_{123}&=&\frac{\pi^3}{8}\Big[W_{123}(u_1,~u_2,~u_3’)+W_{123}(u_1,~u_2’,~u_3)  \nonumber \\
&&+W_{123}(u_1’,~u_2,~u_3)-W_{123}(u_1’,~u_2’,~u_3’)\Big],\nonumber \\
&=&\frac{\pi^2}{4}\Big\{\exp\left[-2|u_3’|^2\right]W_{12}\left(u_1,~u_2\right) \nonumber \\
&&+\exp\left[-2|u_3|^2\right]W_{12}\left(u_1,~u_2’\right) \nonumber \\
&&+\exp\left[-2|u_3|^2\right]W_{12}\left(u_1’,~u_2\right) \nonumber \\
&&-\exp\left[-2|u_3’|^2\right]W_{12}\left(u_1’,~u_2’\right)\Big\}.
\end{eqnarray} 
Taking $u_3=u_3’=0$, we have
\begin{eqnarray}
\mathcal{B}_{123}&=&\frac{\pi^2}{4}\Big[W_{12}\left(u_1,~u_2\right)+W_{12}\left(u_1,~u_2’\right)+W_{12}\left(u_1’,~u_2\right) \nonumber\\
&&-W_{12}\left(u_1’,~u_2’\right)\Big], \nonumber \\
&=&\mathcal{B}_{12}.
\label{Artifical_TriPartiteState_BiTriBell}
\end{eqnarray} 
The above formula indicates that the tripartite Bell function of the artificial three-mode system  is exactly equal to the bipartite Bell function of the 1-2 two-mode squeezed subsystem. We plot the maximized bipartite Bell function of the subsystems and tripartite Bell function of the full system through numerically optimizing parameters $u_j$ and $u_j’~(j=1,2,3)$, as displayed in Fig.~\ref{Fig_ArtificialThreeMode}(c). It is easily seen that the black dashed curve (representing the maximized tripartite Bell function $|\mathcal{B}_{123}|_{\max}$) coincides perfectly with the red solid curve (representing the maximized bipartite Bell function $|\mathcal{B}_{12}|_{\max}$ of the 1-2 subsystem). We note that the tripartite Bell nonlocality is present for any $r>0$, even though the genuine tripartite entanglement is strictly zero. This demonstrates conclusively that genuine tripartite entanglement is not a prerequisite for the violation of tripartite Bell nonlocality.

\bibliography{Bell_nonlocality}
\end{document}